\documentclass[12pt]{iopart}

\usepackage{iopams}  
\usepackage{color}
\usepackage{graphicx}
\begin{document}

\title{The role of second-order radial density gradient for helicon power absorption}
\author{Runlong Wang$^{1}$, Lei Chang$^{1, 2}$, Xinyue Hu$^2$, Lanlan Ping$^3$, Ning Hu$^1$, Xianming Wu$^4$, Jianyao Yao$^1$, Xinfeng Sun$^4$, and Tianping Zhang$^4$}
\address{$^1$ College of Aerospace Engineering, Chongqing University, Chongqing 400044, China}
\address{$^2$ School of Aeronautics and Astronautics, Sichuan University, Chengdu 610065, China}
\address{$^3$ School of Electronic and Information Engineering, Anhui Jianzhu University, Hefei 230601, China}
\address{$^4$ National Key Laboratory of Science and Technology on Vacuum Technology and Physics, Lanzhou Institute of Physics, Lanzhou 730000, China}
\ead{leichang@cqu.edu.cn}
\date{\today}

\begin{abstract}
To reveal the mysterious and still controversial mechanism of high ionization efficiency during helicon discharges, this work focuses particularly on the role of second-order derivative in radial density profile, both analytical and numerically. It is found that: (i) the radially localized potential well that plays a critical role in resonance power absorption from antenna to plasma is localized where the second-order derivative vanishes, (ii) the power absorption increases for positive second-order derivative, decreases for negative second-order derivative, and maximizes where second-order derivative becomes zero, (iii) the power absorption decreases near plasma core and increases near plasma edge when the radial location of vanishing second-order derivative moves outwards, which is also a process of Trivelpiece-Gould mode overwhelming helicon mode. These findings can be very interesting for helicon plasma applications that require certain power distribution or heat flux configuration, e. g. material processing, which can be controlled by adjusting the profile and zero-crossing location of second-order radial density gradient. 
\end{abstract}



\maketitle

\section{Background}
Helicon wave is a bounded whistler mode propagating in magnetized plasma cylinders and with frequency between the ion and electron cyclotron frequencies.\cite{Boswell:1970aa, Boswell:1997aa, Chen:1997aa} It can generate discharge with remarkable plasma density and ionization efficiency, which are much higher than those of inductively and capacitively coupled plasma discharges with similar power input, and therefore attracts great interests from many fields. These include, to name a few, plasma rocket propulsion,\cite{Arefiev:2004aa, Ziemba:2005aa, Charles:2009aa, Batishchev:2009aa} plasma source for magnetic fusion studies,\cite{Loewenhardt:1991aa} Alfv\'{e}n wave propagation,\cite{Hanna:2001aa} radio-frequency current drive,\cite{Petrzilka:1994aa} laser plasma sources,\cite{Zhu:1989aa} semiconductor processing, electrodeless beam sources, and laser accelerators.\cite{Chen:1996aa} However, the underlying mechanism of this surprising ionization efficiency has not yet been fully explained, although two hypotheses are very promising: (i) mode conversion between quasi-electrostatic Trivelpiece-Gould (TG) mode near plasma edge and electromagnetic helicon mode across the whole plasma volume,\cite{Shamrai:1996aa} (ii) radially localized potential well that caused by plasma density gradient (surface-type helicon mode) yielding resonant absorption of radio-frequency power.\cite{Breizman:2000aa} Recently, it was supposed that the existence of positive second-order derivative in the radial density configuration promotes the power absorption near plasma core,\cite{Chang:2016ab} but detailed physical study has been missing. This paper explores this gap and treats the role of second-order radial density gradient both analytically and numerically during the procedure of helicon power absorption. It will show that the power absorption increases for positive second-order derivative, decreases for negative second-order derivative, and maximizes where the second-order derivative vanishes; moreover, the radial location of zero-crossing second-order derivative determines the spatial distribution of power absorption, which can be of great practical interest. The paper is organized as follows: Sec.~\ref{analysis} describes our theoretical analysis from the governing wave equations and based on three-parameter plasma density function, Sec.~\ref{helic} introduces the employed numerical scheme and conditions, Sec.~\ref{profile} and Sec.~\ref{location} present the computed results and physical discussions, and Sec.~\ref{summary} summarizes this work with comments for future research. 

\section{Theoretical analysis}\label{analysis}
In a right-hand cylindrical coordinate system of $(r,~\theta,~z)$ with perturbation in form of $\exp[i(k z+m \theta-\omega t)]$, the wave equations for helicon mode and TG mode propagating in non-uniform plasma can be written in form of wave electric field $E$ as:\cite{Breizman:2000aa}
\small
\begin{equation}\label{eq1}
\frac{1}{r}\frac{\partial}{\partial r}\left[r\frac{\partial E_*}{\partial r}\right]-\frac{m^2}{r^2}E_*=-\frac{m}{k^2 r}\frac{\omega^2}{c^2}\frac{E_*\partial g/\partial r}{1+(m\partial g/\partial r)/k^2 r \eta},
\end{equation}
\begin{equation}\label{eq2}
\frac{1}{r}\frac{\partial}{\partial r}\left[\varepsilon r\frac{\partial E_z}{\partial r}\right]-\frac{m}{r}\left[\frac{\partial g}{\partial r}+\frac{\varepsilon m}{r}\right] E_z-k^2\eta E_z=0, 
\end{equation}
\normalsize
respectively, with $\omega$ the wave frequency, $m$ the azimuthal mode number, $k$ the axial mode number, and $c=1/\sqrt{\epsilon_0\mu_0}$ the speed of light. Here, $E_*$ is a new function $E_*=E_z-(k r/m)E_\theta$ measuring the radial component of perturbed magnetic field through $H_r=(m c/\omega r)E_*$, and can be replaced by an algebraic relationship
\small
\begin{equation}\label{eq3}
E_*=E_z\left[1+k^2 r \eta \left(m\frac{\partial g}{\partial r}\right)^{-1}\right]
\end{equation}
\normalsize
for $\varepsilon\ll g$ and sufficiently smooth radial scale of $E_z-E_*$. Parameters $\varepsilon$, $g$ and $\eta$ are the components of standard cold-plasma dielectric tensor, and can be expressed in forms of: 
\small
\begin{equation}\label{eq4}
\varepsilon=1-\sum_\alpha\frac{\omega_{p\alpha}^2}{\omega^2-\omega_{c\alpha}^2},~g=-\sum_\alpha\frac{\omega_{c\alpha}\omega_{p\alpha}^2}{\omega(\omega^2-\omega_{c\alpha}^2)},~\eta=1-\sum_\alpha\frac{\omega_{p\alpha}^2}{\omega^2}
\end{equation}
\normalsize
with the subscript $\alpha$ labelling particle species (ion and electron), $\omega_{c\alpha}=q_\alpha B_0/m_\alpha$ the cyclotron frequency, and $\omega_{p\alpha}=\sqrt{n_\alpha(r) q_\alpha^2/\epsilon_0 m_\alpha}$ the plasma frequency. We can see that all these three components depend on the zero-order plasma density $n_\alpha(r)$, and since $\varepsilon\ll g$ for most helicon discharges the radial density gradient that effects the mode structure and eventually the power absorption is mainly through $\partial g/\partial r$ term in Eqs.~(\ref{eq1}-\ref{eq3}). As stated previously, the radial density gradient involves a surface-type mode, which is actually a new helicon mode coupled to space charge effects from electron $\mathbf{E}\times\mathbf{B}$ drift in non-uniform plasma.\cite{Breizman:2000aa} The corresponding perturbed ``surface" current is localized around the peak of $E_*(r)$ and distinguishes the power absorption mechanism from other hypotheses.\cite{Breizman:2000aa, Klozenberg:1965aa, Sudan:1967aa} Please note that the maximum of $E_*(r)$ corresponds to $n''(r)=0$ for smooth density distributions, therefore, the second-order derivative of radial density profile deserves particular attention as being focused in the present work. 

\begin{figure}[ht]
\begin{center}$
\begin{array}{ll}
(a)&(b)\\
\includegraphics[width=0.51\textwidth,angle=0]{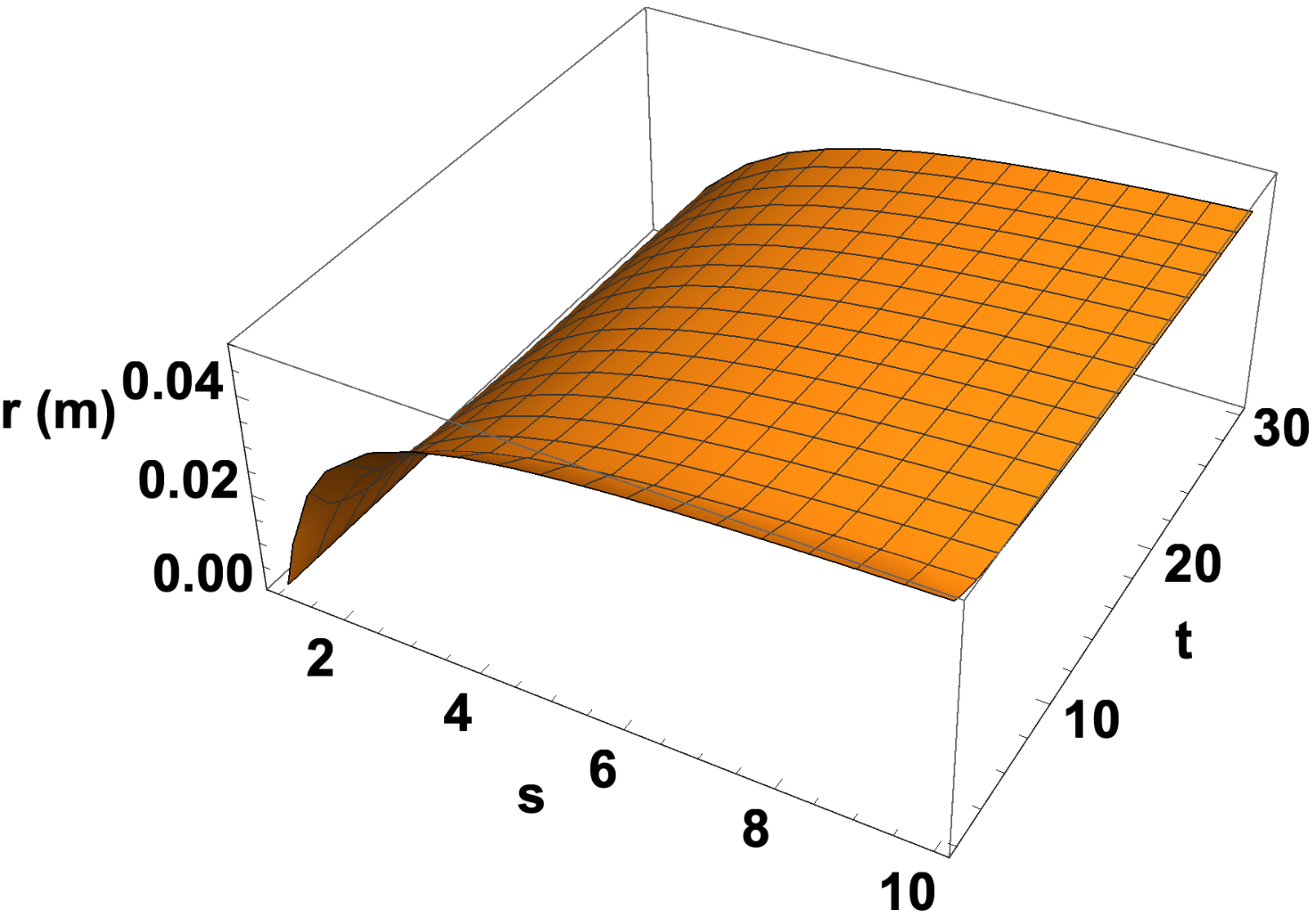}&\includegraphics[width=0.46\textwidth,angle=0]{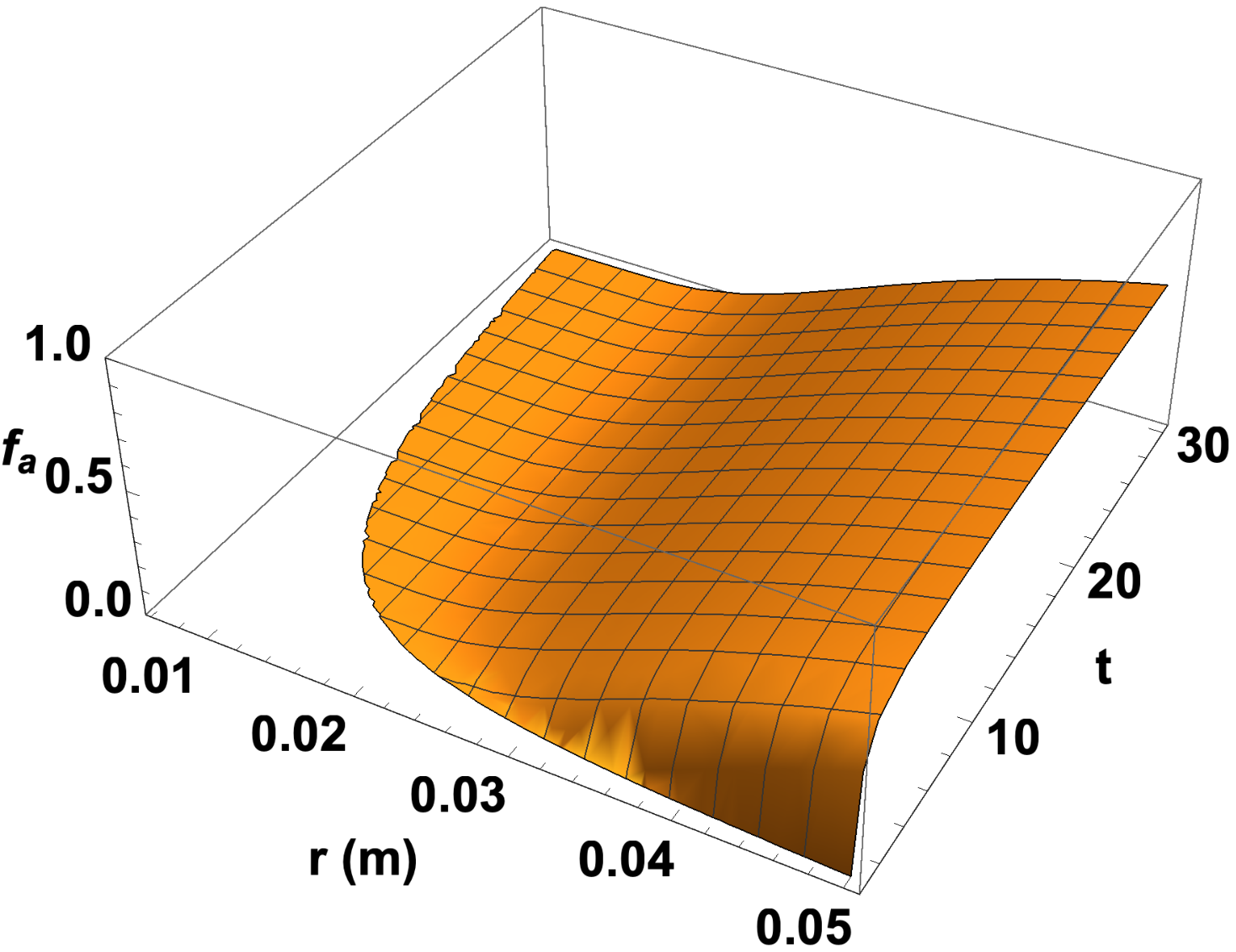}
\end{array}$
\end{center}
\caption{Visualization of the constraint $(s t-1)(1-f_a^{1/t})\left(\frac{r}{a}\right)^{s}-(s-1)=0$ that yields $n_*''(r)=0$: (a) dependence of $r$ on $s$ and $t$ for $f_a=0.01$ and $a=0.05$~m, (b) dependence of $f_a$ on $r$ and $t$ for $s=2$ and $a=0.05$~m.}
\label{fg1}
\end{figure}

To calculate the second-order radial density gradient, we consider a three-parameter $(f_a,~s,~t)$ function of the form
\small
\begin{equation}\label{eq9}
\frac{n(r)}{n_0}=n_*(r)=\left[1-\left(\frac{r}{\xi}\right)^s\right]^t,~\xi=\frac{a}{(1-f_a^{1/t})^{1/s}},
\end{equation}
\normalsize
which could describe various helicon plasma configurations such as uniform, parabolic, linear and Gaussian. Please note that this three-parameter function is incorporated into the HELIC code that will be employed in the following sections. Here, $n_0$ labels the on-axis density, $n_*(r)$ represents the normalized density profile, $a$ is the plasma radius, $f_a$ stands for the normalized density at edge $n_*(a)$, and $s$ and $t$ are adjusting parameters. It is then straightforward to derive the first-order and second-order derivatives of $n_*(r)$:
\small
\begin{equation}\label{eq10}
n_*'(r)=-\frac{s t}{\xi}\left(\frac{r}{\xi}\right)^{s-1}\left[1-\left(\frac{r}{\xi}\right)^s\right]^{t-1},
\end{equation}
\begin{equation}\label{eq11}
n_*''(r)=\frac{s t}{\xi^2}\left(\frac{r}{\xi}\right)^{s-2}\left[1-\left(\frac{r}{\xi}\right)^s\right]^{t-2}\left[(s t-1)\left(\frac{r}{\xi}\right)^{s}-(s-1)\right].
\end{equation}
\normalsize
For positive $s$ and $t$, the sign of $n_*''(r)$ is determined by the sign of last square bracket, which can be then set to zero with the replacement of $\xi$
\small
\begin{equation}\label{eq12}
(s t-1)(1-f_a^{1/t})\left(\frac{r}{a}\right)^{s}-(s-1)=0
\end{equation}
\normalsize
to find constraint relations in the space of $(r;~f_a,~s,~t)$ for $n_*''(r)=0$. We will show later that $n_*''(r)=0$ is correlated with the maximum power absorption, and the radial location of $n_*''(r)=0$ determines the radial power distribution. Therefore, this constraint can be very useful for adjusting $(r;~f_a,~s,~t)$ to find required $n_*''(r)=0$. As an illustration, we visualize the constraint in Fig.~\ref{fg1}, which is also a reference for following sections. 

\section{Numerical computation}\label{helic}
We utilize the well-known HELIC code to compute the wave field and power absorption for different plasma configurations. HELIC is a C++ program developed by Donald Arnush and Francis Chen in the 1990s for designing efficient radio-frequency plasma sources and interpreting experimental data.\cite{Chen:1997ab, Arnush:1997aa, Arnush:2000aa} The underlying theoretical model consists of Maxwell's equations, which can be manipulated into four coupled first-order differential equations for the Fourier transformed variables:\cite{Arnush:2000aa}
\small
\begin{equation}\label{eq13}
\frac{\partial E_\theta}{\partial r}=\frac{i m}{r}E_r-\frac{E_\theta}{r}+i\omega B_z,
\end{equation}
\begin{equation}\label{eq14}
\frac{\partial E_z}{\partial r}=i k E_r-i\omega B_\theta,
\end{equation}
\begin{equation}\label{eq15}
i\frac{\partial B_\theta}{\partial r}=\frac{m}{r}\frac{k}{\omega}E_\theta-\frac{i B_\theta}{r}+\left(\eta_*-\frac{m^2}{k_0^2 r^2}\right)\frac{\omega}{c^2}E_z,
\end{equation}
\begin{equation}\label{eq16}
i\frac{\partial B_z}{\partial r}=\frac{\omega}{c^2}i g_* E_r+(k^2-k_0^2 \varepsilon_*)\frac{E_\theta}{\omega}+\frac{m}{r}\frac{k}{\omega}E_z. 
\end{equation}
\normalsize
Here, $B$ is wave magnetic field, $k_0=\omega/c$ and a phenomenological collision frequency $\nu$ has been introduced to the dielectric elements to resolve the electron-neutral and electron-ion collisions: 
\small
\begin{equation}\label{eq17}
\varepsilon_*=1-\sum_\alpha\frac{\omega+i\nu_\alpha}{\omega}\frac{\omega_{p\alpha}^2}{(\omega+i\nu_\alpha)^2-\omega_{c\alpha}^2},
\end{equation}
\begin{equation}\label{eq18}
g_*=-\sum_\alpha\frac{\omega_{c\alpha}}{\omega}\frac{\omega_{p\alpha}^2}{(\omega+i\nu_\alpha)^2-\omega_{c\alpha}^2},
\end{equation}
\begin{equation}\label{eq19}
\eta_*=1-\sum_\alpha\frac{\omega_{p\alpha}^2}{\omega(\omega+i\nu_\alpha)}.
\end{equation}
\normalsize
This collision frequency depends on electron temperature $T_e$ and varies with radius for non-uniform radial density profiles. In principle, the HELIC code can include any number of ion species and displacement current, although a single species of singly charged ions is considered throughout this paper. Through a standard subroutine, the code solves $E$ and $B$ from Eqs.~(\ref{eq13}-\ref{eq16}) for different values of $k$, while the $m$ is fixed for each time of running, and we can further compute the current density $\mathbf{j}$ and power absorption $P(r,~z)$ based on the solved $E$ and $B$. 

\begin{figure}[ht]
\begin{center}
\includegraphics[width=0.55\textwidth,angle=0]{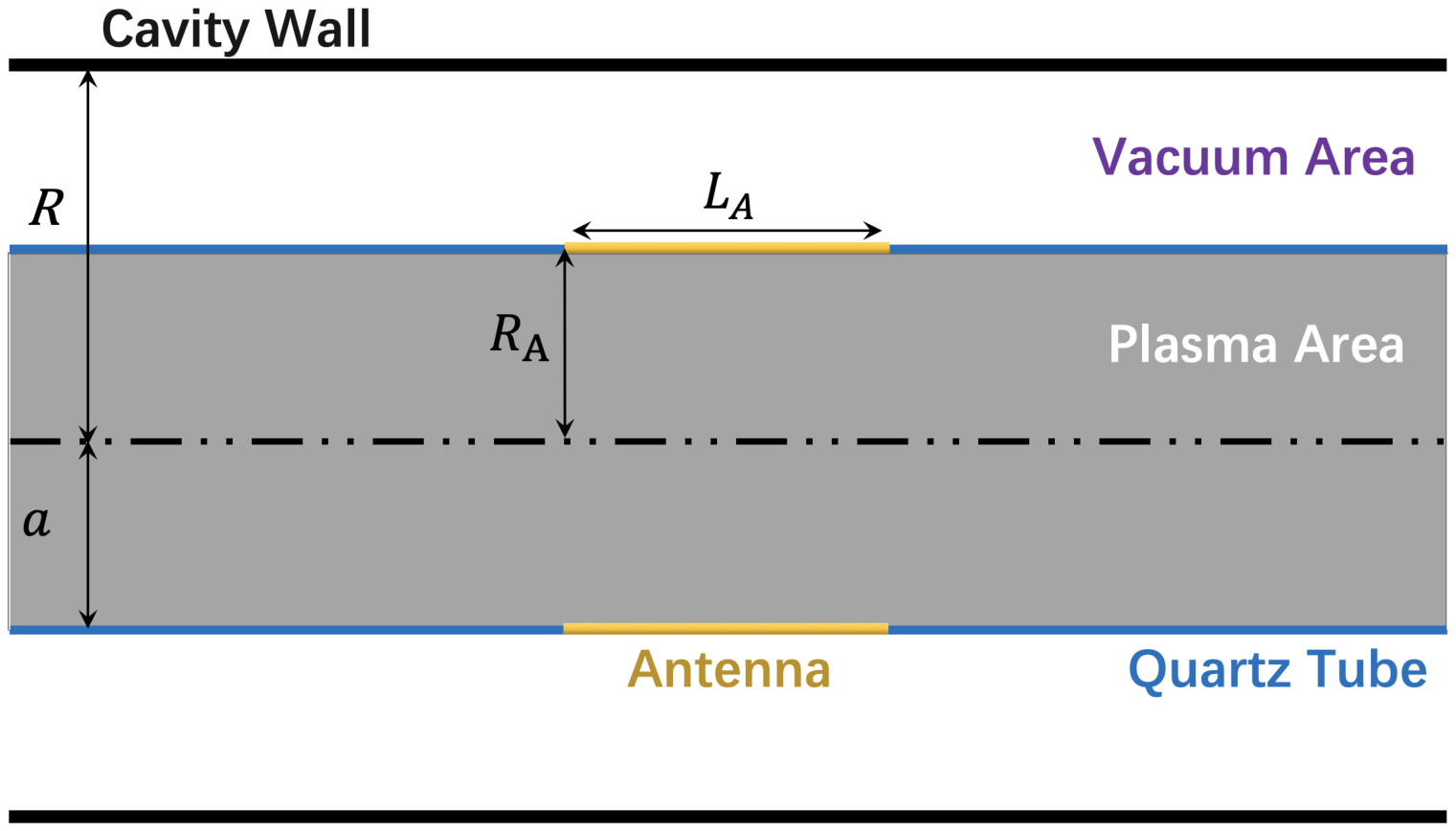}
\end{center}
\caption{A schematic of the employed computational domain.}
\label{fg2}
\end{figure}

A schematic of the computational domain is drawn in Fig.~\ref{fg2}. The formed plasma is contained by a cylindrical quartz tube, which is insulated from the outer cavity wall by vacuum area. The radius of the cavity wall is $R$ where the tangential components of wave electric field vanishes due to conducting boundary conditions assumed. A half-turn helical antenna, driven by radio-frequency power supply, wraps around the middle of quartz tube and sustains the plasma discharge inside. The length and radius of this antenna are $L_A$ and $R_A$, respectively, and the thickness is assumed to be zero so that only a current sheet is considered in the HELIC computations. To eliminate the reflection from axial endplates, the domain is assumed to be infinitely long, although the actual length of $-0.63\leq z\leq 0.63$~(m) is considered, and computations are limited to the spectral range of $5\leq k\leq 50$~($1/\textrm{m}$) for the conditions employed. 
Specific parameters are listed in Tab.~\ref{tab1}, which are typical measurements in various helicon discharge experiments.\cite{Boswell:1984ab, Boswell:1987aa, Blackwell:2012aa, Squire:2006aa, Mori:2004aa} We assume that the plasma has no axial structure, to focus on the effect of radial density gradient on helicon power absorption, and only considers the original Spitzer collision frequency  for simplicity. 
\begin{table}
\caption{\label{tab1} Employed parameters for HELIC computation.}
\footnotesize
\hspace{3cm}\begin{tabular}{ll}
Parameters & Values\\
\hline
Plasma radius ($a$) & $0.05$~m\\
Antenna radius ($R_A$) & $0.055$~m\\
Antenna length ($L_A$) & $0.2$~m\\
Cavity radius ($R$) & $0.15$~m\\
Cavity length & Axially unbounded\\
Plasma density ($n_0$) & $10^{18}~\textrm{m}^{-3}$\\
Antenna type & Half helix ($m$ odd)\\
Antenna frequency ($\omega$) & $13.56$~MHz\\
Azimuthal mode ($m$) & 1\\
Static magnetic field ($B_0$) & $0.02$~T\\
Ion species & Ar\\
Electron temperature ($T_e$)& $3$~eV\\
Background pressure ($P_0$)& $1.33$~Pa
\end{tabular}
\normalsize
\end{table}

\section{Radial profile of $n_*''(r)$}\label{profile}
\begin{figure}[ht]
\begin{center}$
\begin{array}{ll}
(a)&(b)\\
\includegraphics[width=0.465\textwidth,angle=0]{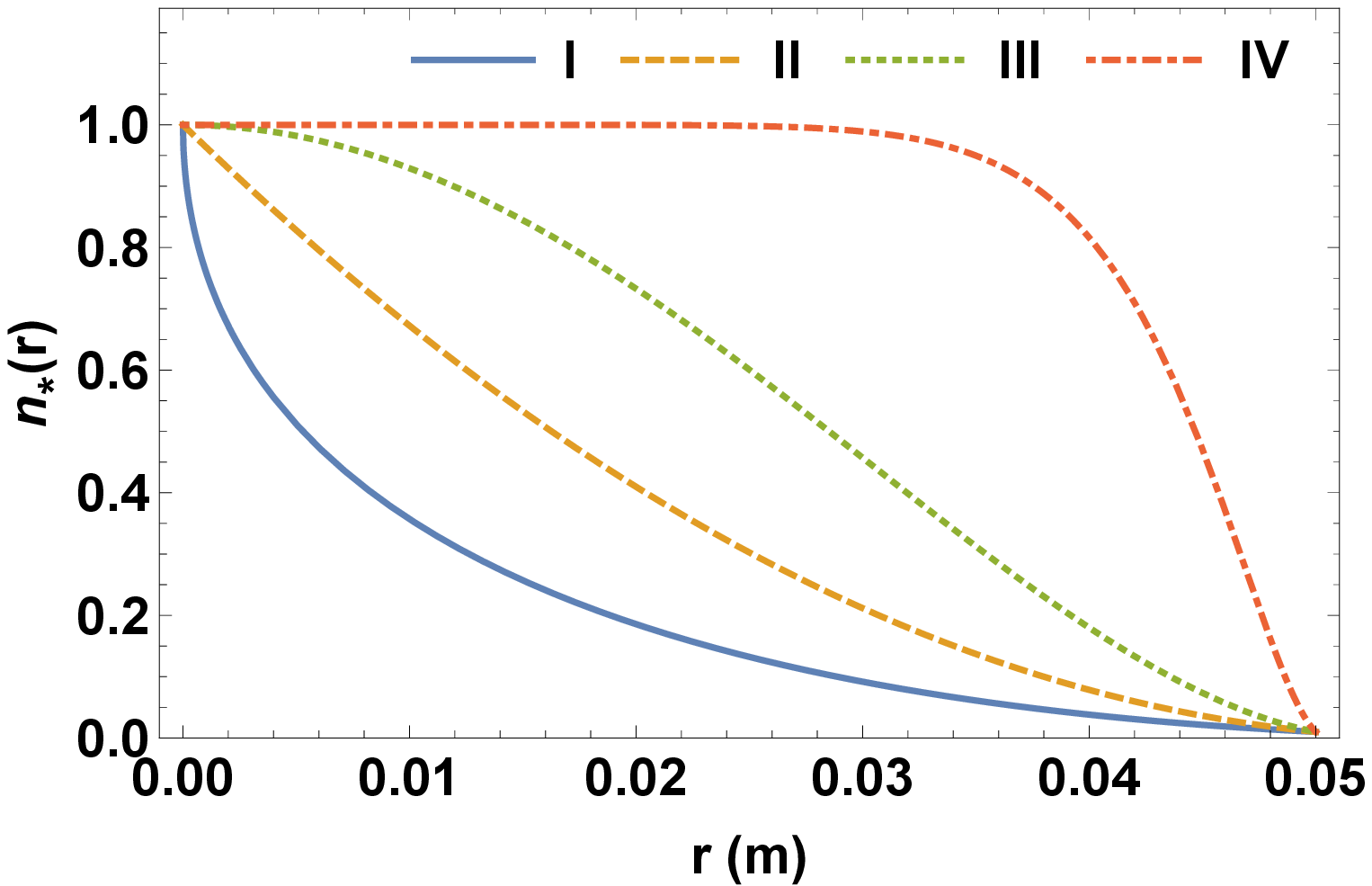}&\includegraphics[width=0.51\textwidth,angle=0]{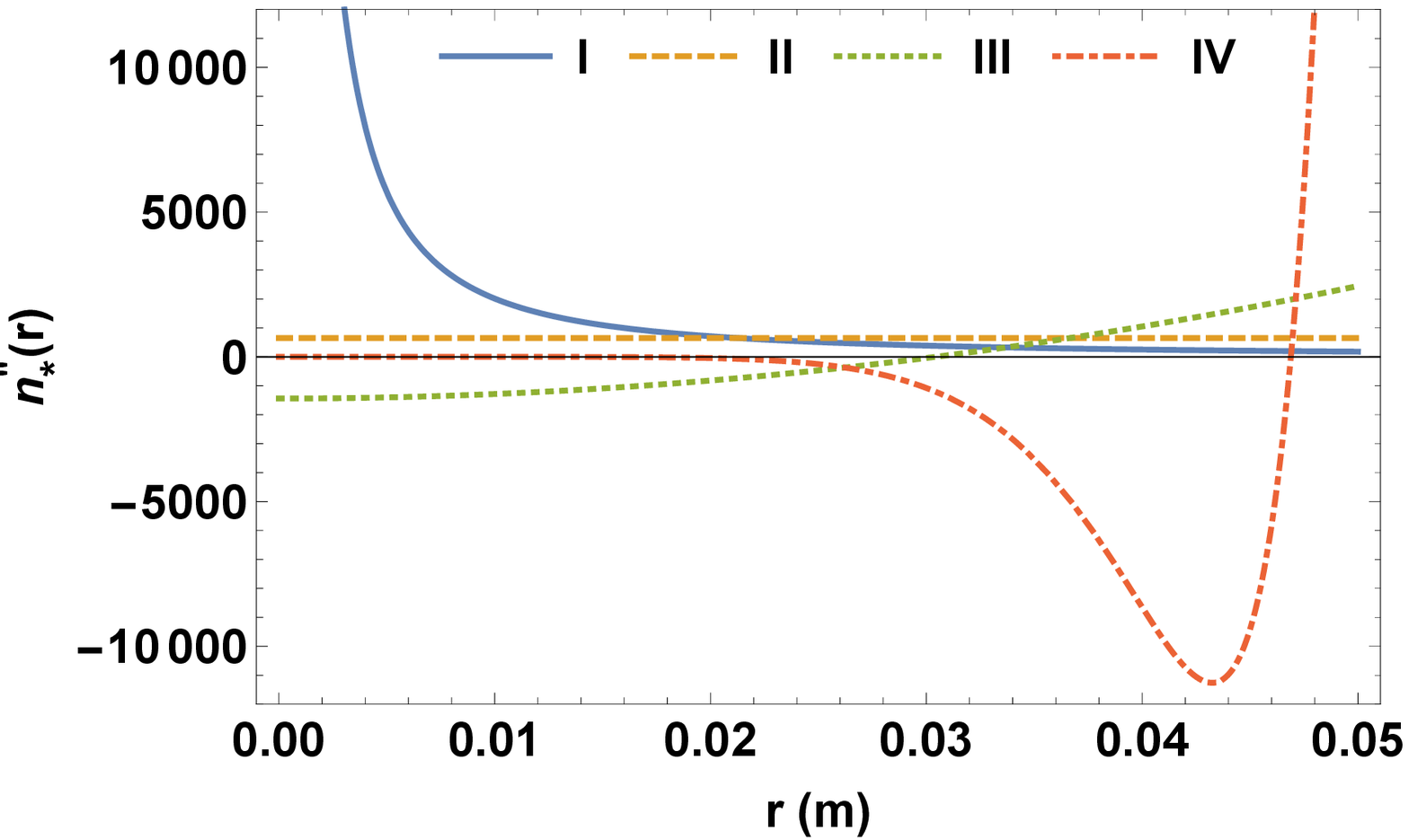}
\end{array}$
\end{center}
\caption{Radial profiles of normalized plasma density $n_*(r)$ and its second-order derivative $n_*''(r)$ constructed for HELIC computations: I ($f_a=0.01,~s=0.5,~t=2$), II ($f_a=0.01,~s=1,~t=2$), III ($f_a=0.01,~s=2,~t=2$), IV ($f_a=0.01,~s=10,~t=2$). }
\label{fg3}
\end{figure}

\begin{figure}[ht]
\begin{center}$
\begin{array}{ll}
(a)&(b)\\
\includegraphics[width=0.485\textwidth,angle=0]{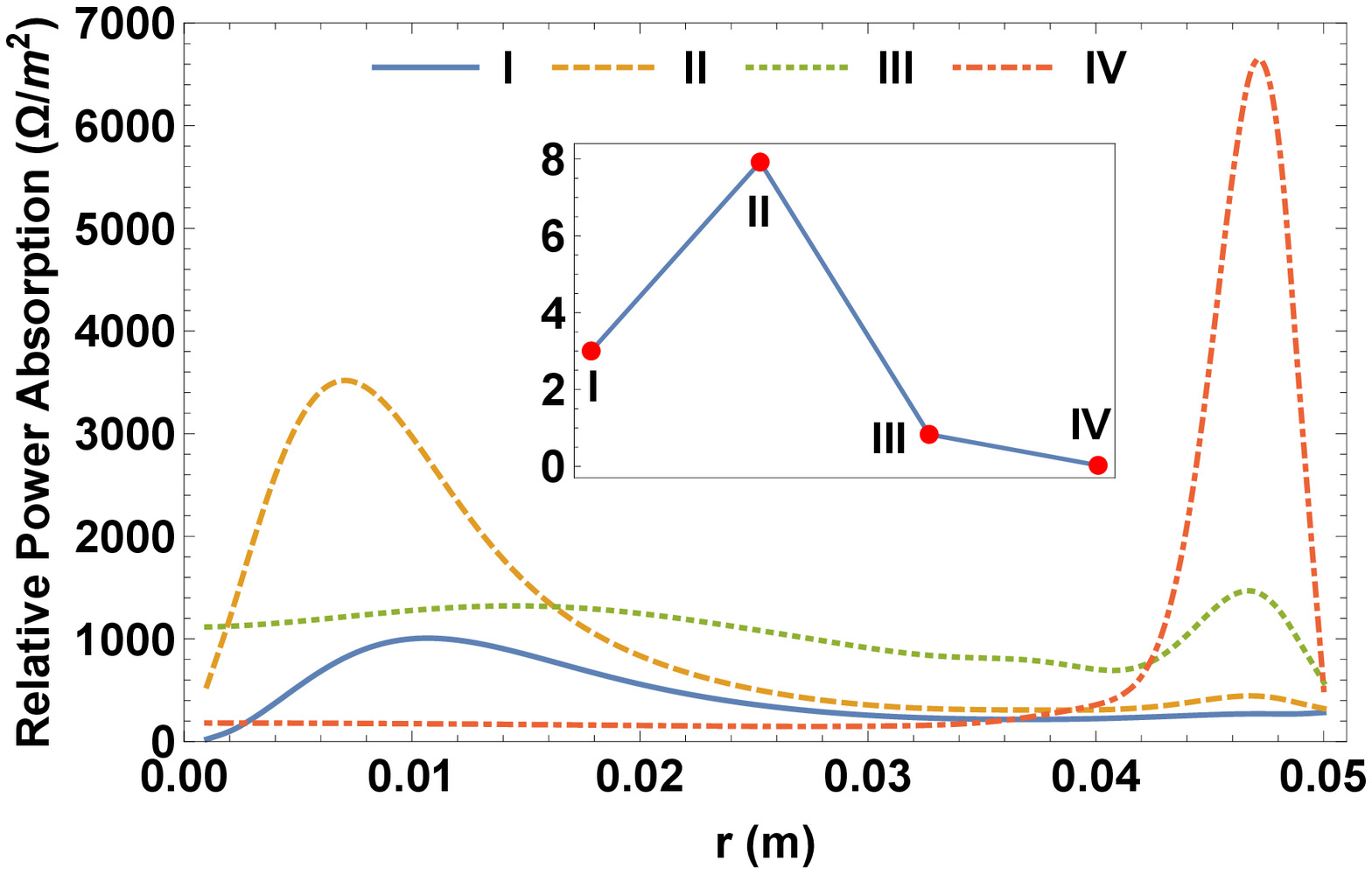}&\includegraphics[width=0.464\textwidth,angle=0]{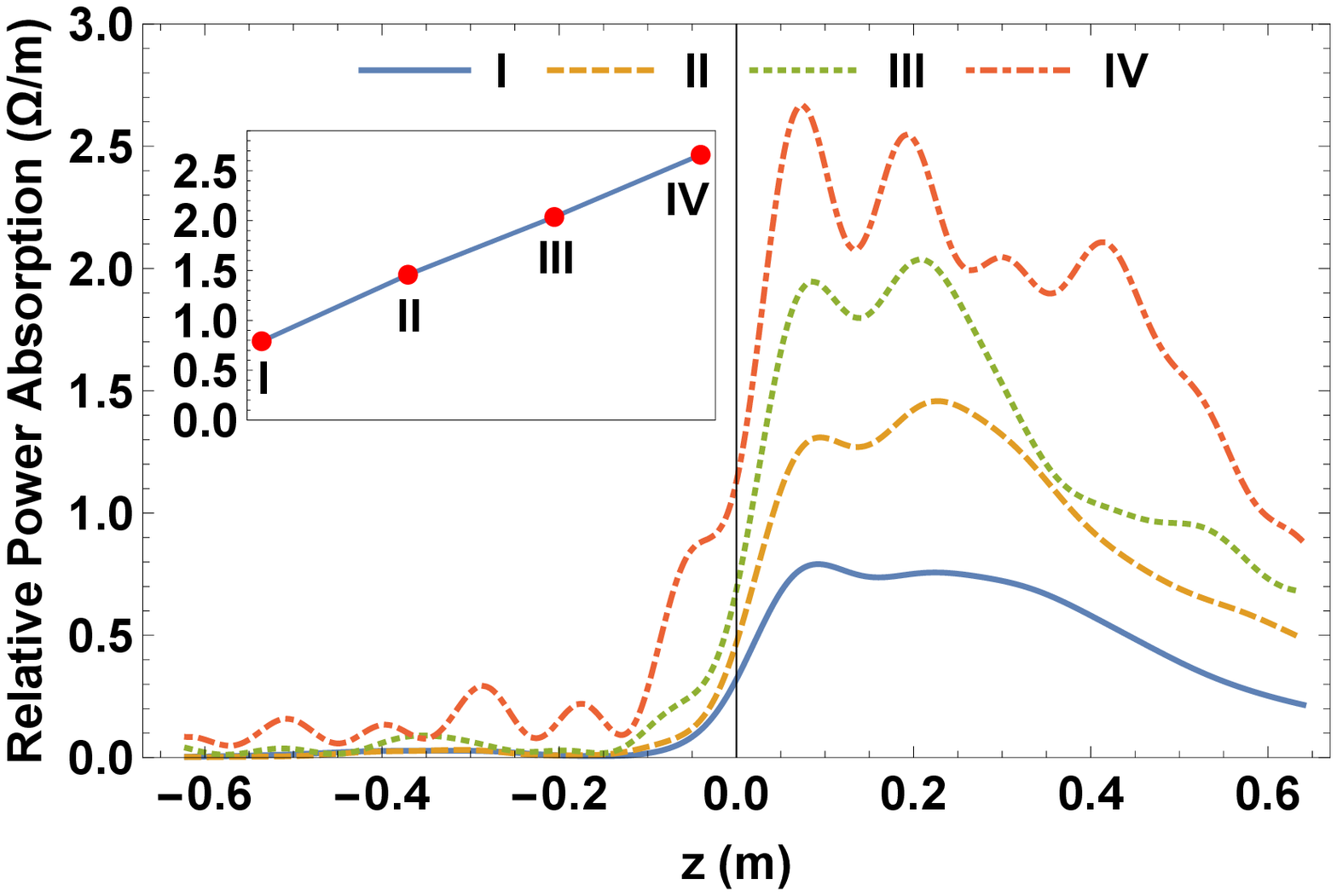}
\end{array}$
\end{center}
\caption{Relative power absorption in: (a) radial ($z=0.2$~m), (b) axial ($r=0.02$~m) directions. The inset figures show: (a) the ratio of $P_r(0.007)/P_r(0.047)$, (b) the peak magnitude of $P_z(z)$ for different density profiles constructed in Fig.~\ref{fg3}.}
\label{fg4}
\end{figure}

Figure~\ref{fg3} shows the employed radial profiles of plasma density, which are constructed based on the three-parameter function given in Eq.~\ref{eq9}, and the corresponding second-order derivatives. These profiles are typical for present study: Profile-I ($f_a=0.01,~s=0.5,~t=2$) shows positive and monotonously decreasing $n_*''(r)$, Profile-II ($f_a=0.01,~s=1,~t=2$) displays positive and constant $n_*''(r)$, Profile-III ($f_a=0.01,~s=2,~t=2$) illustrates monotonously increasing $n_*''(r)$ from negative to positive, and Profile-IV ($f_a=0.01,~s=10,~t=2$) represents non-monotonic and mostly negative $n_*''(r)$. 
The computed relative power absorptions in radial ($z=0.2$~m) and axial ($r=0.02$~m) directions are shown in Fig.~\ref{fg4}. Please note here that: $P_r$ is the relative absorption at each radius, integrated from $Z_{min}$ to $Z_{max}$; $P_z$ is the relative absorption at each $z$, integrated over the plasma cross section; the ``relative" phrase refers to the antenna current of $1$ A. The ratio of core ($r=0.007$~m) to edge ($r=0.047$~m) power absorption in radial direction and the peak magnitude of power absorption in axial direction are also plotted as inset figures. We can see that, as the second-order derivative of radial density profile descends from positive to negative values, the ratio of power absorption near core to that near edge largely decreases, although the positive and constant second-order derivative seems yield a maximum ratio. This is consistent with a previous study claiming that the existence of positive second-order derivative in radial density profile results in more power absorbed near plasma core.\cite{Chang:2016ab} Moreover, the peak and total power absorption (computed by integrating $\mathbf{E}\cdot \mathbf{J}$ over the plasma volume)  increases. This implies that the core heating is less dominant than edge heating in the sense of global power absorption, or alternatively edge heating is necessary for high-efficiency helicon discharge. 

\begin{figure}[ht]
\begin{center}$
\begin{array}{ll}
(a)&(b)\\
\includegraphics[width=0.485\textwidth,angle=0]{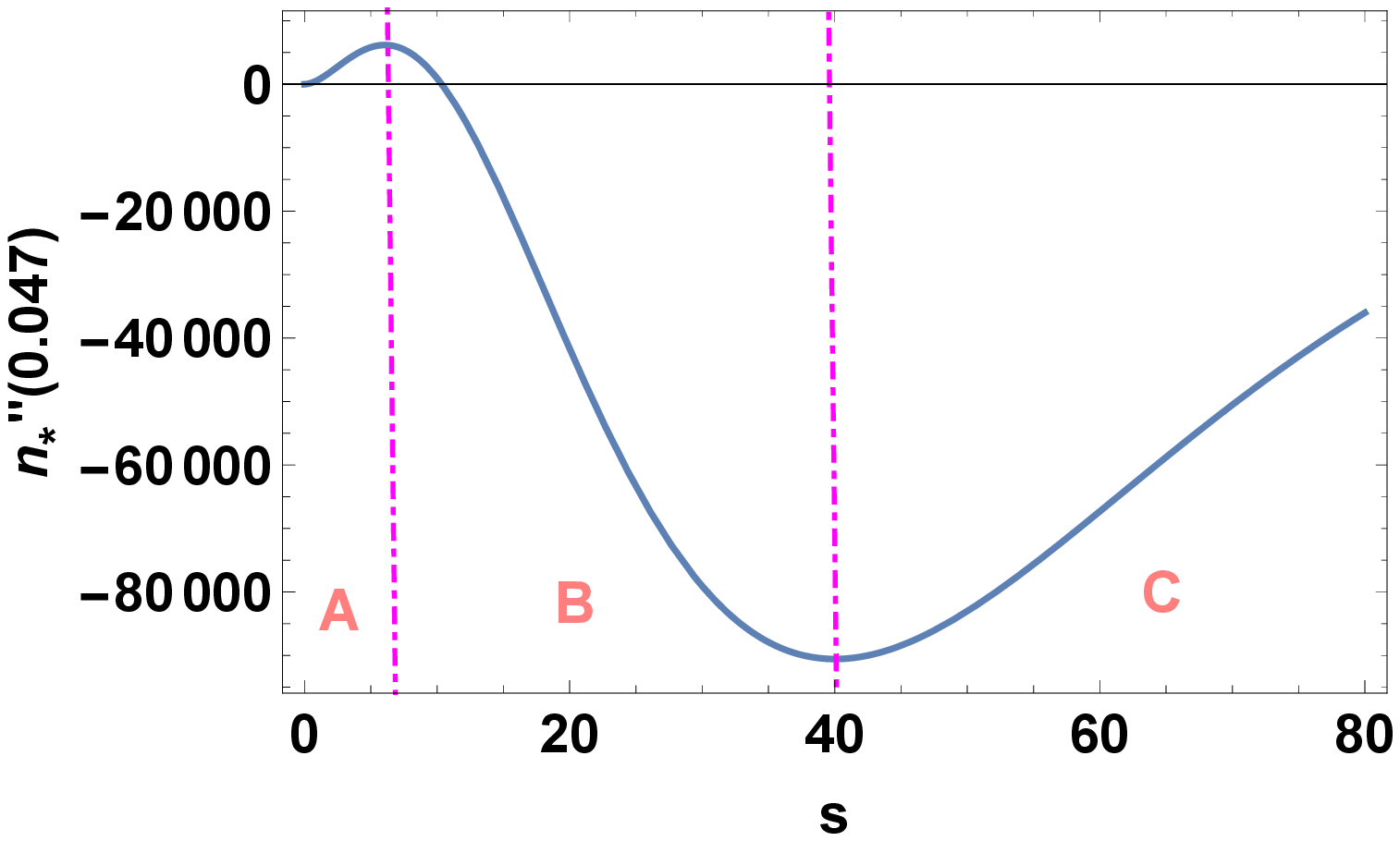}&\includegraphics[width=0.46\textwidth,angle=0]{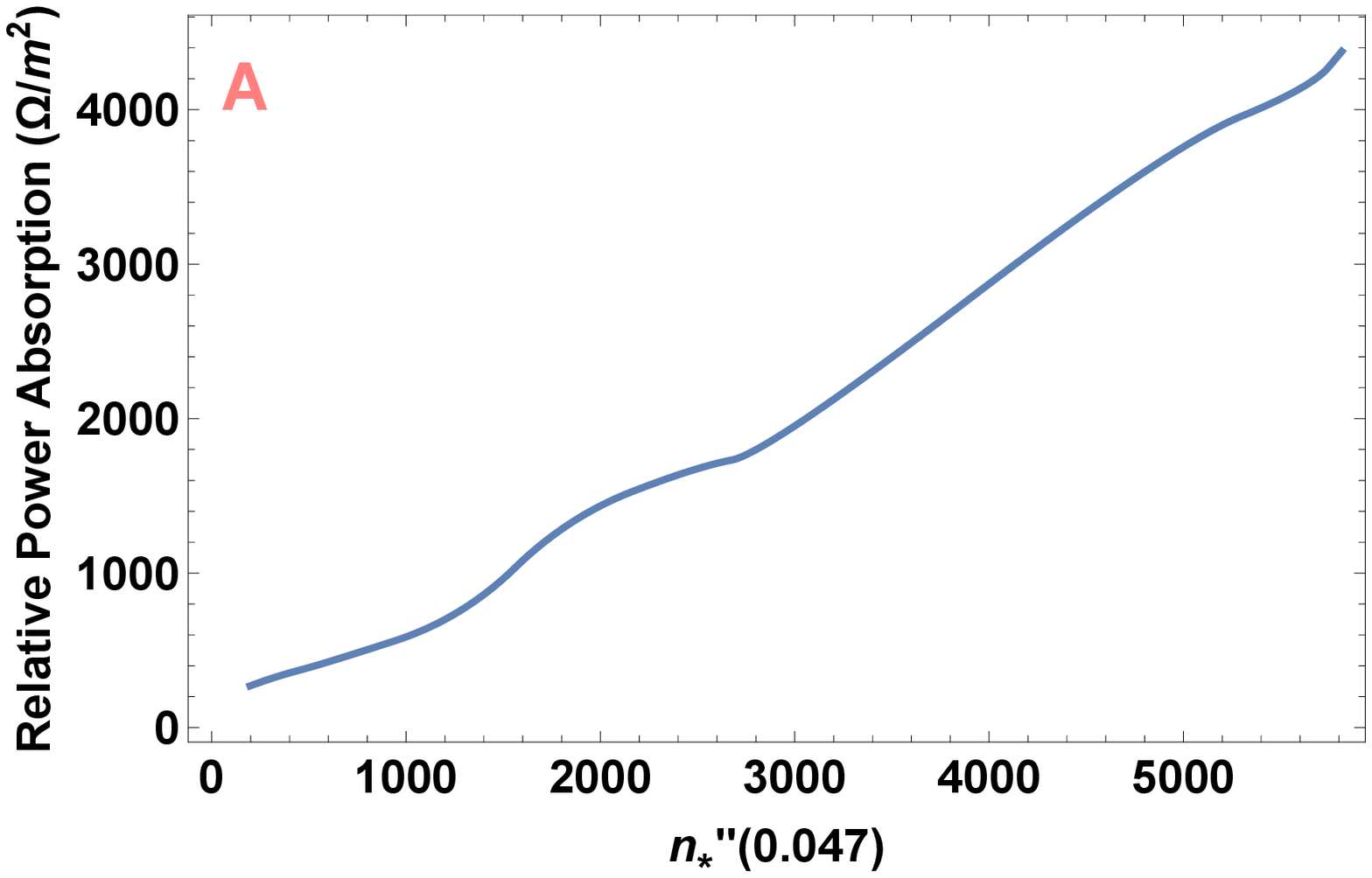}\\
(c)&(d)\\
\hspace{0.55cm}\includegraphics[width=0.45\textwidth,angle=0]{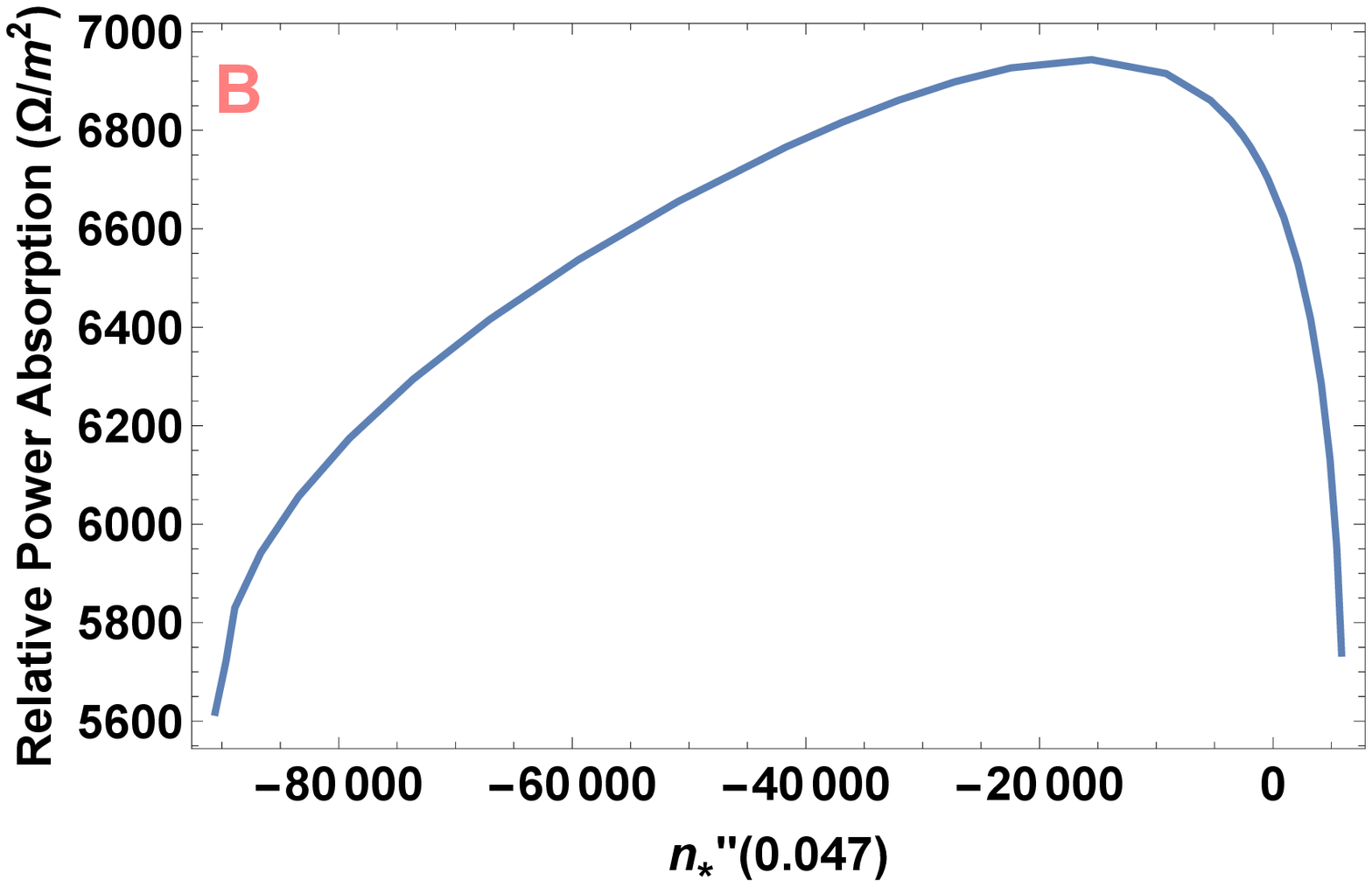}&\includegraphics[width=0.46\textwidth,angle=0]{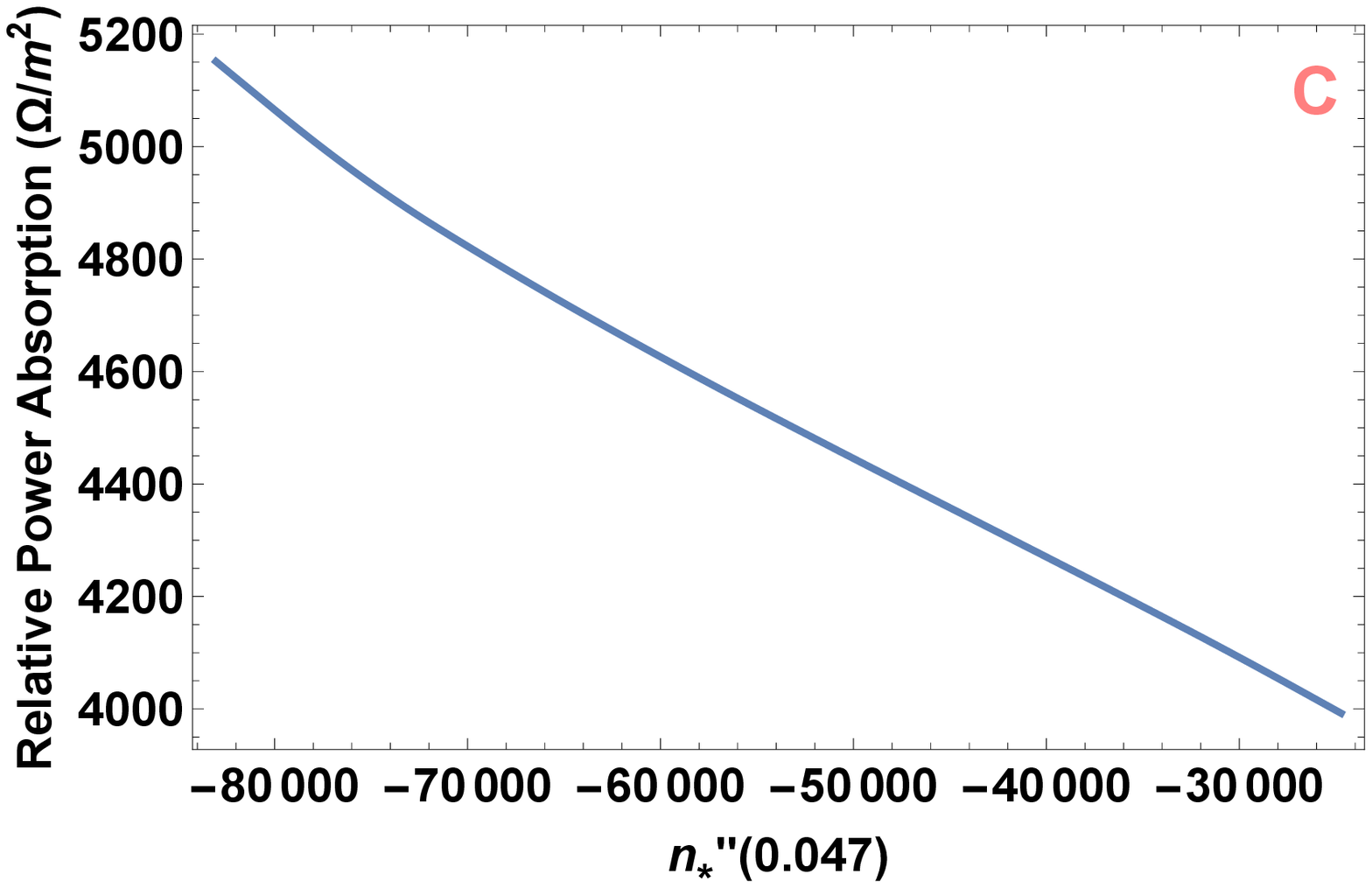}
\end{array}$
\end{center}
\caption{Dependence of $n_*''(0.047)$ on $s$ (a) and the variation of relative power absorption with $n_*''(0.047)$ (b-d) in three different areas of (a): (A) monotonously increasing and always positive, (B) monotonously decreasing from positive to negative, (C) monotonously increasing but always negative.}
\label{fg6}
\end{figure}

Since the radial power absorption maximizes near $r=0.007$~m and $r=0.047$~m, as shown in Fig.~\ref{fg4}(a), we scan the second-order density gradient at these two radial locations as a function of $s$, while $f_a=0.01$ and $t=2$ are remained, and further divide the plot into three areas: (A) monotonously increasing and always positive, (B) monotonously decreasing from positive to negative, (C) monotonously increasing but always negative. 
\begin{figure}[ht]
\begin{center}$
\begin{array}{ll}
(a)&(b)\\
\includegraphics[width=0.48\textwidth,angle=0]{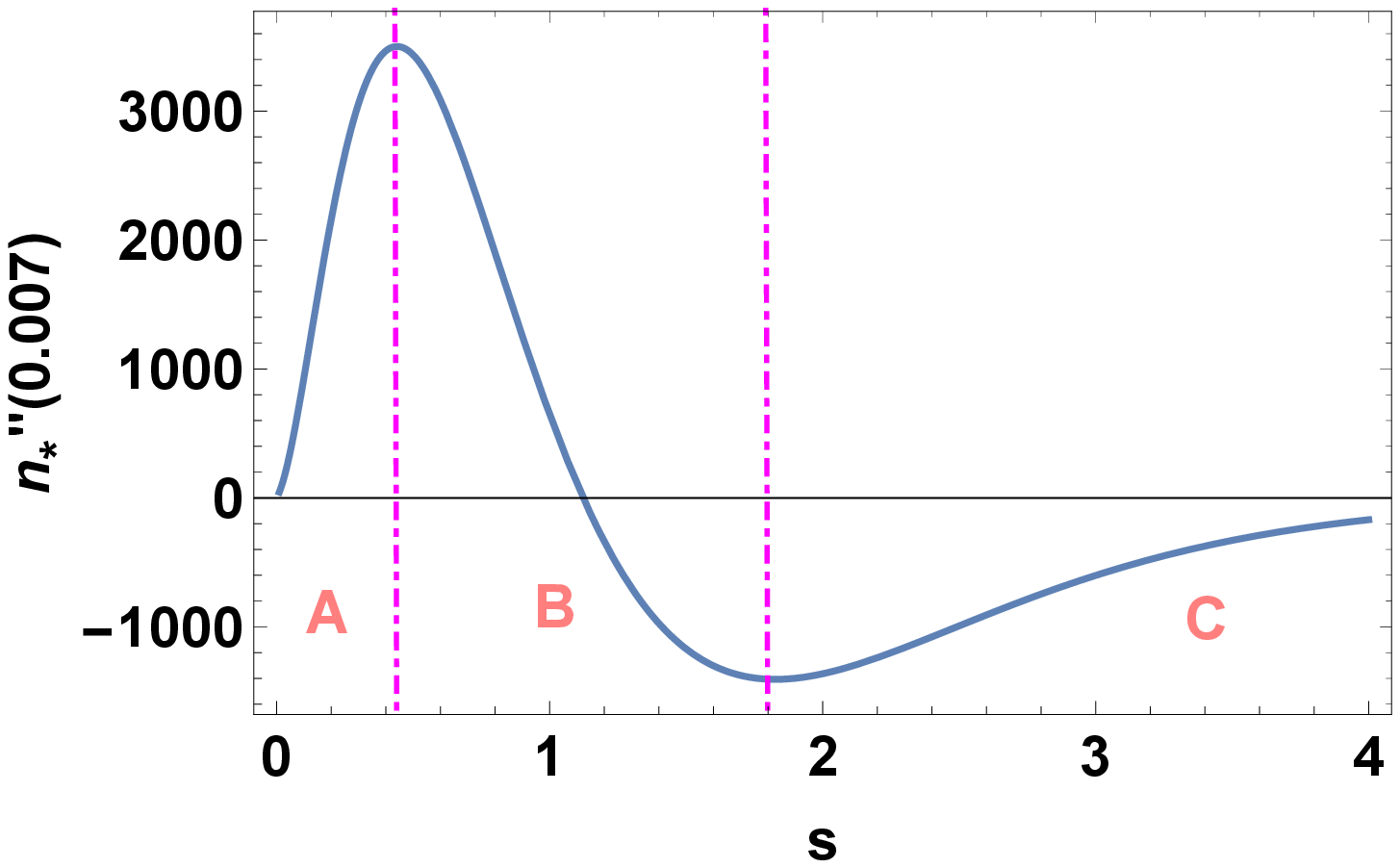}&\includegraphics[width=0.478\textwidth,angle=0]{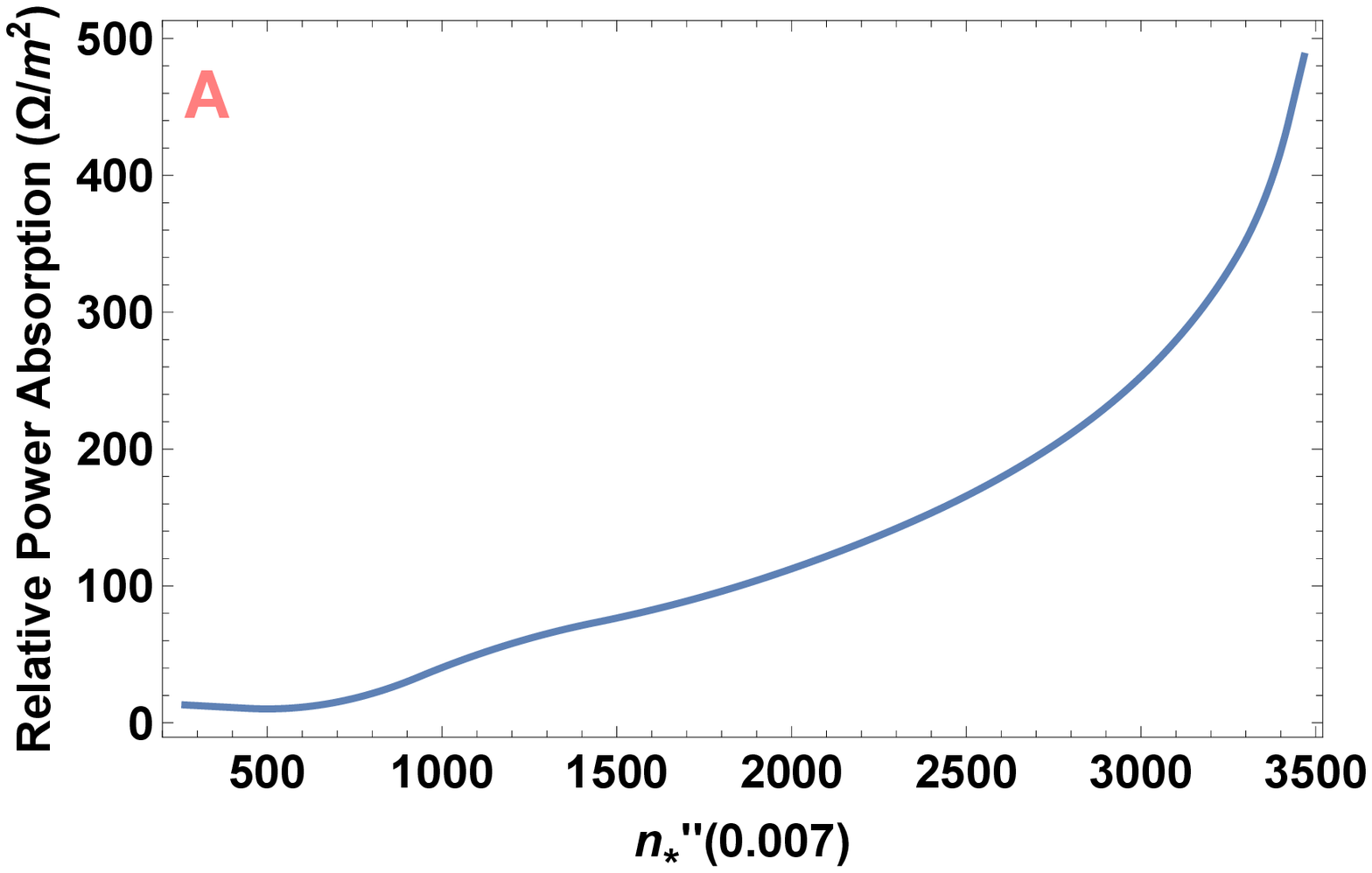}\\
(c)&(d)\\
\hspace{0.35cm}\includegraphics[width=0.46\textwidth,angle=0]{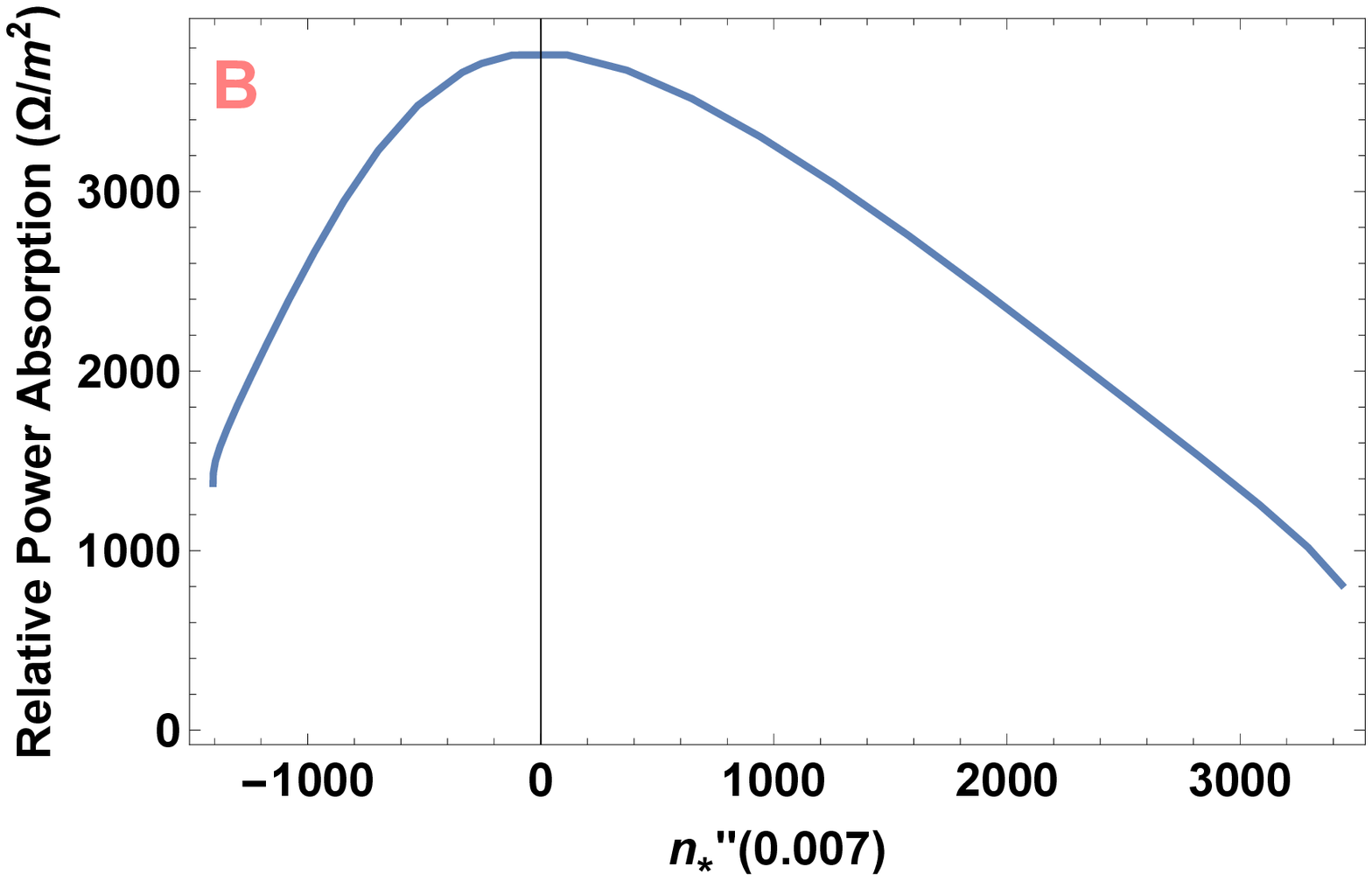}&\includegraphics[width=0.465\textwidth,angle=0]{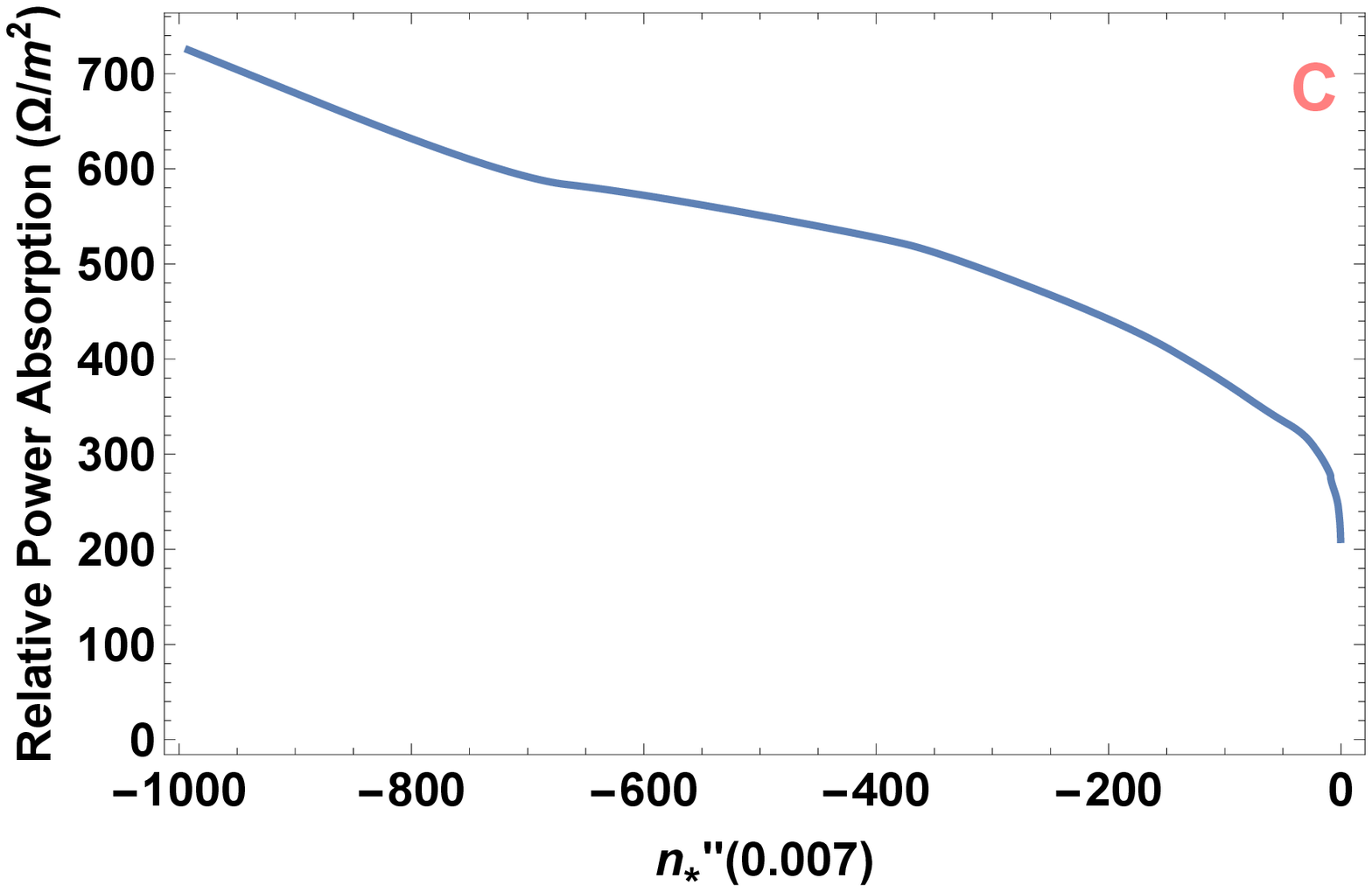}
\end{array}$
\end{center}
\caption{Dependence of $n_*''(0.007)$ on $s$ (a) and the variation of relative power absorption with $n_*''(0.007)$ (b-d) in three different areas of (a): (A) monotonously increasing and always positive, (B) monotonously decreasing from positive to negative, (C) monotonously increasing but always negative.}
\label{fg5}
\end{figure}
Then we visualize the dependence of relative power absorption as a function of $n_*''(0.007)$ and $n_*''(0.047)$ in these three areas. As shown in Fig.~\ref{fg5} and Fig.~\ref{fg6}, respectively, the power absorption increases with positive and growing $n_*''(0.007)$ and $n_*''(0.047)$ (A-area), and decreases with negative and growing $n_*''(0.007)$ and $n_*''(0.047)$ (C-area). Interestingly, for the second-order density gradient descending from positive to negative values (B-area), the power absorption increases with positive and dropping $n_*''(0.007)$ and $n_*''(0.047)$ till a maximum and then decreases with negative and dropping $n_*''(0.007)$ and $n_*''(0.047)$. The maximum power absorptions occur near $n_*''(0.007)=0$ and $n_*''(0.047)=0$, respectively. Please note that the $x$-coordinates in Fig.~\ref{fg5}(c) and Fig.~\ref{fg6}(c) are from right to left referring to the $x$-coordinates in Fig.~\ref{fg5}(a) and Fig.~\ref{fg6}(a). But overall the relative power absorption increases for positive $n_*''(r)$ whereas decreases for negative $n_*''(r)$, regardless of $n_*''(r)$ growing or dropping, and it maximizes near $n_*''(r)=0$. This is consistent with our analysis in Sec.~\ref{analysis} that the perturbed ``surface" current is localized around the peak of $E_*(r)$ where $n_*''(r)=0$ so that the power absorption also maximizes there. 
Similarly, we make use of $f_a=0.01$ and $s=2$ and vary $t$. The trend of relative power absorption growing with positive $n_*''(r)$ but dropping with negative $n_*''(r)$ is observed again, together with the maximum power absorption occurred near $n_*''(r)=0$. This means that the sign of second-order derivative in radial density profile is indeed very important for helicon power absorption, especially the zero-crossing point. 

\section{Radial location of $n_*''(r)=0$}\label{location}
\begin{figure}[ht]
\begin{center}$
\begin{array}{ll}
(a_1)&(a_2)\\
\includegraphics[width=0.475\textwidth,angle=0]{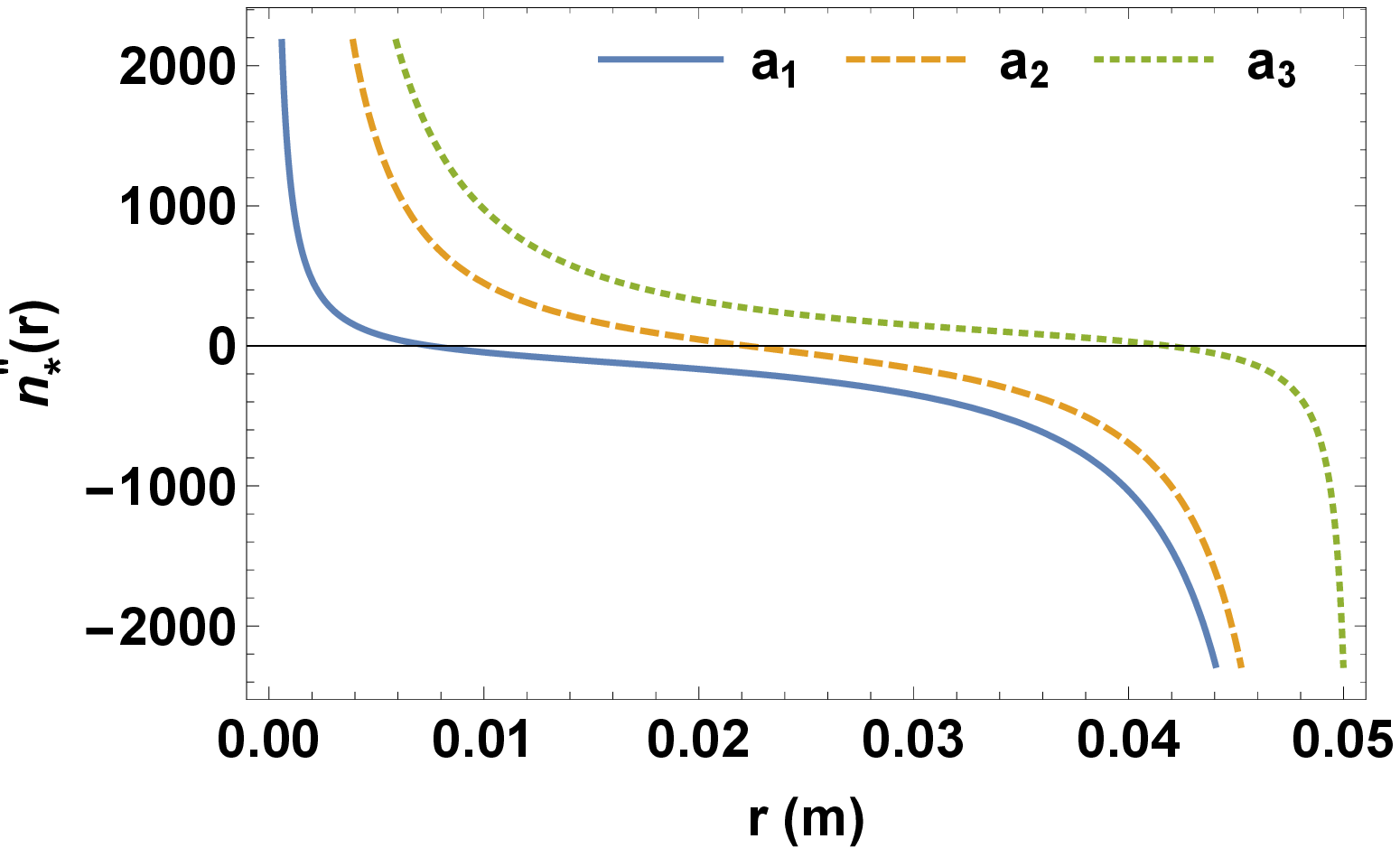}&\includegraphics[width=0.46\textwidth,angle=0]{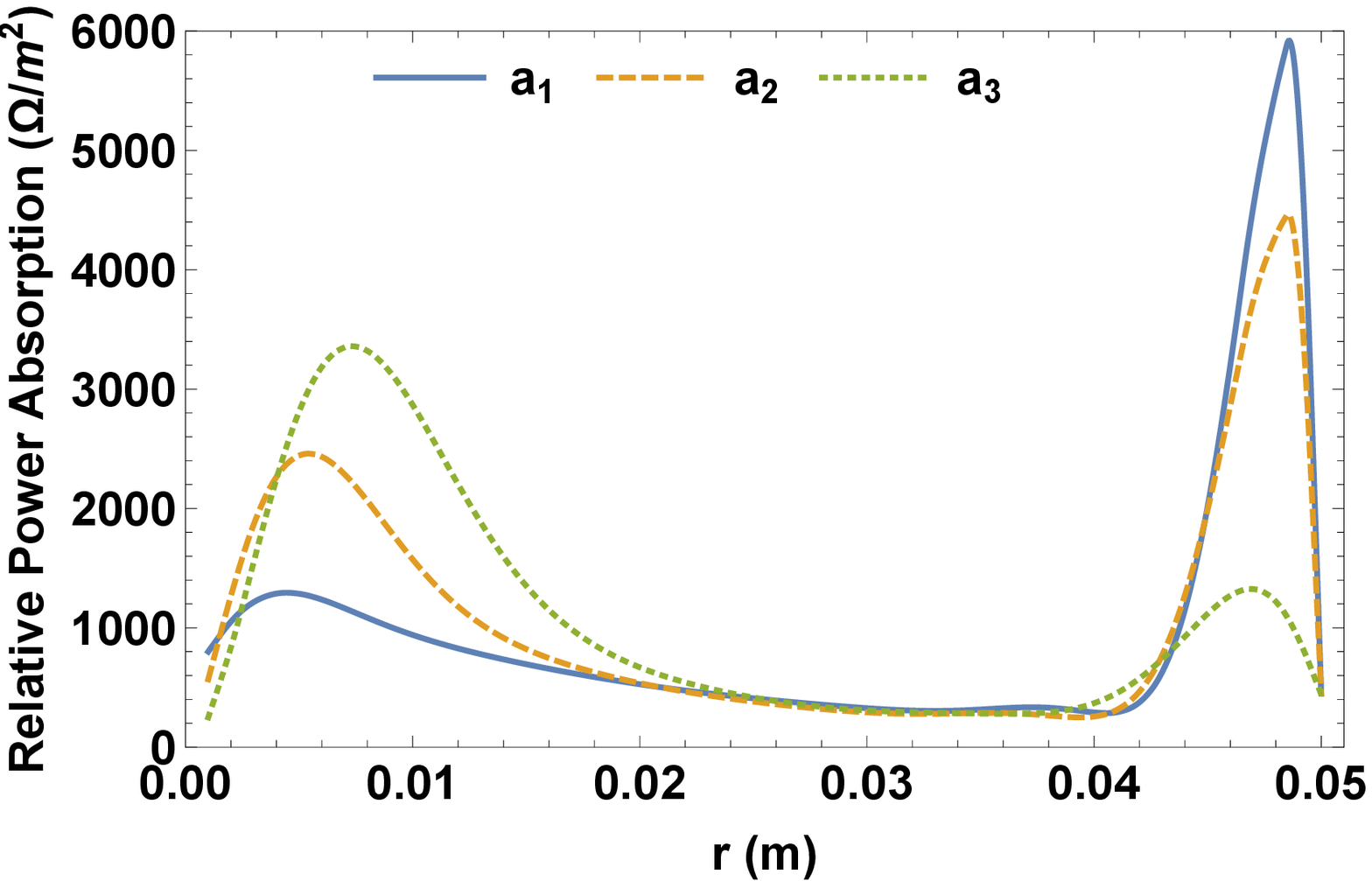}\\
(b_1)&(b_2)\\
\includegraphics[width=0.475\textwidth,angle=0]{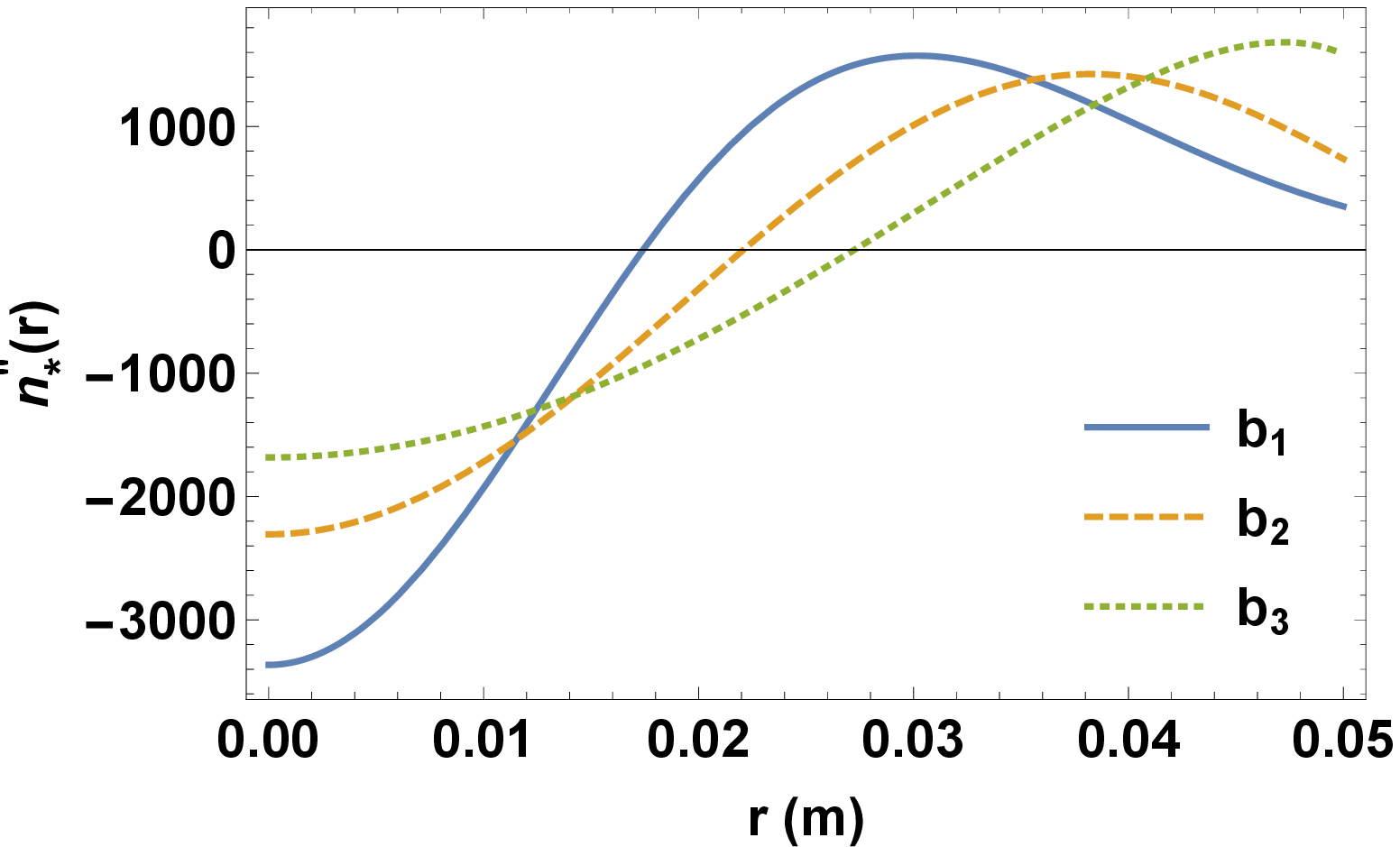}&\includegraphics[width=0.46\textwidth,angle=0]{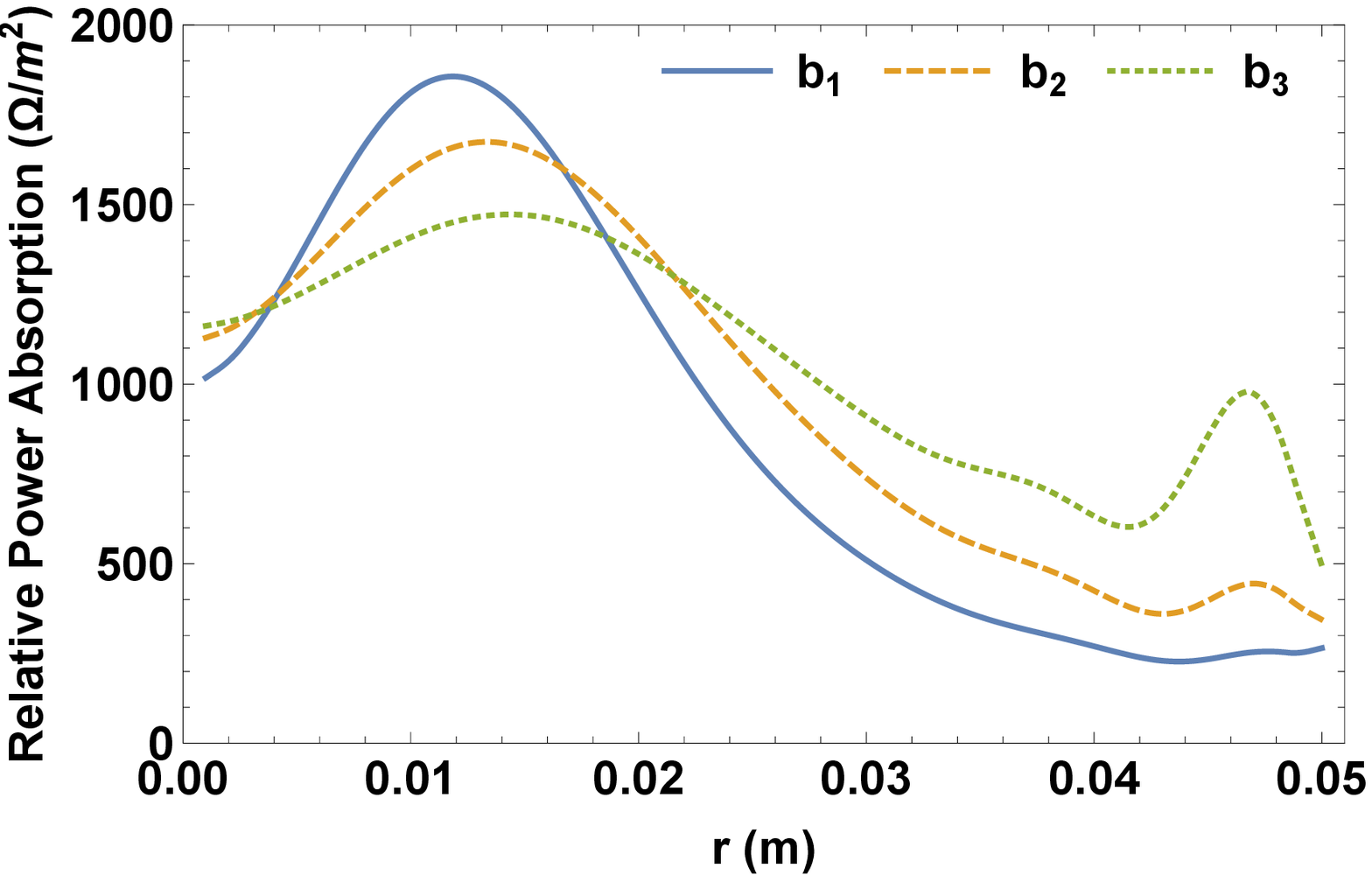}\\
(c_1)&(c_2)\\
\includegraphics[width=0.475\textwidth,angle=0]{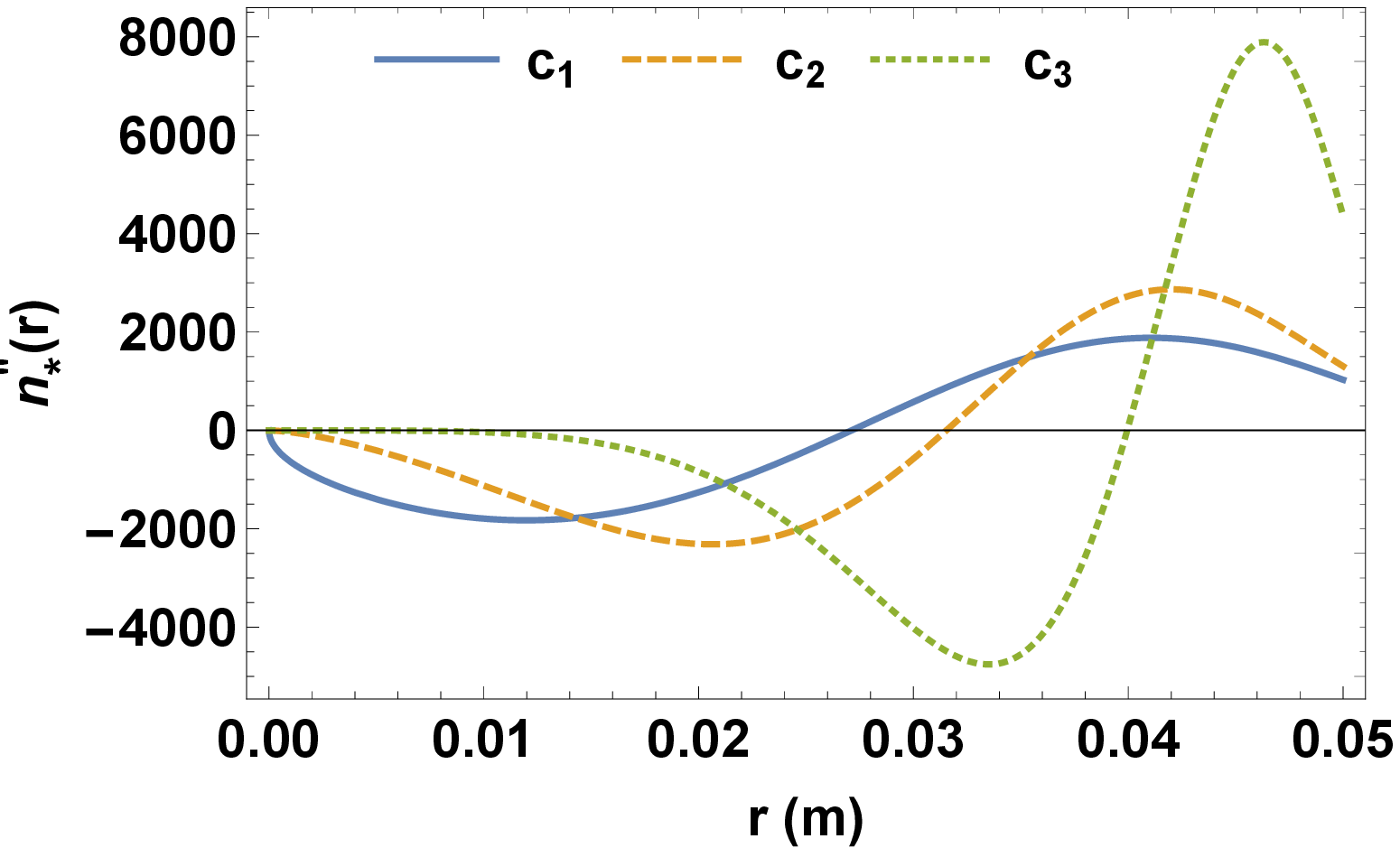}&\includegraphics[width=0.46\textwidth,angle=0]{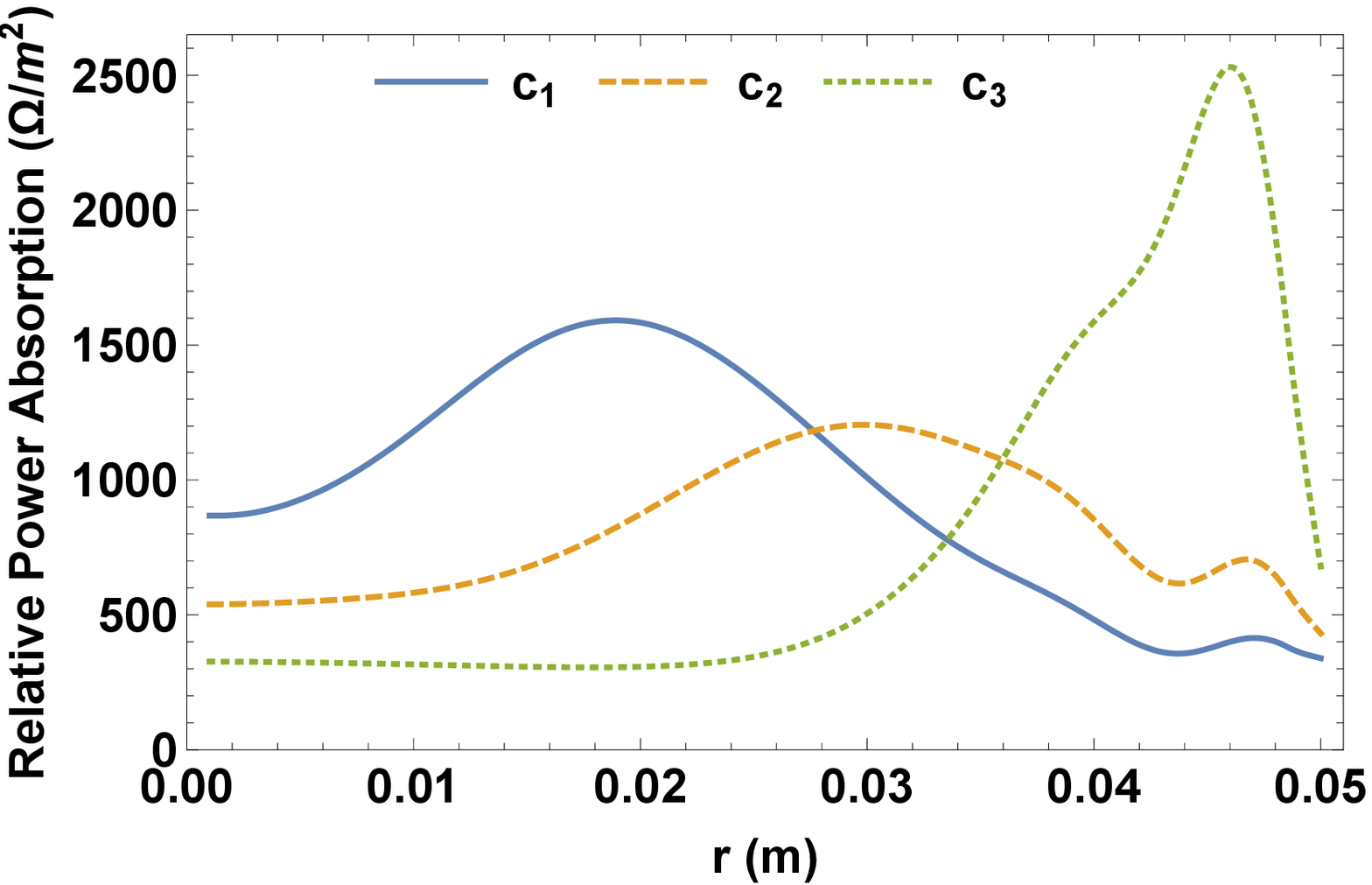}
\end{array}$
\end{center}
\caption{Variations of the radial location of $n_*''(r)=0$ (left) and the corresponding radial profiles of power absorption (right): $a_1$ for $(s, t)=(0.9, 0.5)$, $a_2$ for $(s, t)=(0.5, 0.5)$, $a_3$ for $(s, t)=(0.5, 0.9)$; $b_1$ for $(s, t)=(2, 25)$, $b_2$ for $(s, t)=(2, 4.5)$, $b_3$ for $(s, t)=(2, 2.5)$; $c_1$ for $(s, t)=(2.5, 10)$, $c_2$ for $(s, t)=(3.5, 10)$, $c_3$ for $(s, t)=(6.5, 10)$, with $f_a$ fixed to $0.01$. }
\label{fg7}
\end{figure}

Given that the power absorption maximizes near $n_*''(r)=0$, it is then straightforward and necessary to study the influence of radial location of $n_*''(r)=0$ on the radial power distribution. By varying the density-control parameters of $s$, $t$ and $f_a$, we found there mainly exist three types of radial density configurations that have zero-crossing point of $n_*''(r)=0$. These typical profiles of $n_*''(r)$ are shown in the left figures of Fig.~\ref{fg7}: $a_1$ for $(s, t)=(0.9, 0.5)$, $a_2$ for $(s, t)=(0.5, 0.5)$, $a_3$ for $(s, t)=(0.5, 0.9)$; $b_1$ for $(s, t)=(2, 25)$, $b_2$ for $(s, t)=(2, 4.5)$, $b_3$ for $(s, t)=(2, 2.5)$; $c_1$ for $(s, t)=(2.5, 10)$, $c_2$ for $(s, t)=(3.5, 10)$, $c_3$ for $(s, t)=(6.5, 10)$, with $f_a$ fixed to $0.01$.
The corresponding power absorption in radial direction is illustrated by the right figures of Fig.~\ref{fg7}. We can see that for all these density profiles constructed, as the radial location of $n_*''(r)=0$ moves outwards, the power absorption decreases near plasma core but increases near plasma edge, which also implies that the TG mode becomes stronger than helicon mode during this trend. This can be experimentally very important in the sense that the radial power absorption can be controlled by adjusting the radial location of $n_*''(r)=0$, and is of practical interest for applying helicon source to plasma-material processing, for example, which requires certain distribution of heat flux. 

\section{Summary}\label{summary}
This work follows a previous finding that the existence of positive second-order derivative in radial density profile can promote the radio-frequency power absorption near plasma core, and studies the role of second-order radial density gradient detailedly during helicon power absorption. Through theoretical analysis, we found that the perturbed ``surface" current, which plays a critical role in resonance power absorption from antenna to plasma, is localized where the second-order radial density gradient vanishes, and obtained a constraint relation to find this vanishing gradient based on a three-parameter density function. Employing the well-benchmarked HELIC code, we computed the relative power absorption in both radial and axial directions for typical second-order density gradients in radius: positive and monotonously decreasing, positive and constant, monotonously increasing from negative to positive, and non-monotonic and mostly negative. It is found that the relative power absorption increases for positive second-order derivative, decreases for negative second-order derivative, and maximizes where second-order derivative becomes zero. Moreover, we studied the effect of radial location of vanishing second-order derivative on the radial power distribution, and found that the power absorption decreases near plasma core and increases near plasma edge when this radial location moves outwards, a process of TG mode overwhelming helicon mode. This finding is very useful for certain plasma applications which require particular power distribution or heat flux configuration that can be achieved by adjusting the radial location of vanishing second-order density gradient. Further research can be devoted to experimentally studying the role of second-order radial density gradient in helicon power absorption for comparison. The required profile and zero-crossing point of this gradient may be accomplished by external magnetic field and certain antenna geometry.  

\ack
This work is supported by various funding sources: National Natural Science Foundation of China (11405271), China Postdoctoral Science Foundation (2017M612901), Chongqing Science and Technology Commission (cstc2017jcyjAX0047), Chongqing Postdoctoral Special Foundation (Xm2017109), Fundamental Research Funds for Central Universities (YJ201796), Pre-research of Key Laboratory Fund for Equipment (61422070306), and Shanghai Engineering Research Center of Space Engine (17DZ2280800).

\section*{References}
\bibliographystyle{unsrt}

\begin{thebibliography}{10}

\bibitem{Boswell:1970aa}
R.~W. Boswell.
\newblock Plasma production using a standing helicon wave.
\newblock {\em Physics Letters A}, 33(7):457, 1970.

\bibitem{Boswell:1997aa}
R.~W. Boswell and F.~F. Chen.
\newblock Helicons-the early years.
\newblock {\em Plasma Science, IEEE Transactions on}, 25(6):1229, 1997.

\bibitem{Chen:1997aa}
F.~F. Chen and R.~W. Boswell.
\newblock Helicons-the past decade.
\newblock {\em Plasma Science, IEEE Transactions on}, 25(6):1245, 1997.

\bibitem{Arefiev:2004aa}
A.~V. Arefiev and B.~N. Breizman.
\newblock Theoretical components of the vasimr plasma propulsion concept.
\newblock {\em Physics of Plasmas}, 11(5):2942--2949, 2004.

\bibitem{Ziemba:2005aa}
T.~Ziemba, J.~Carscadden, J.~Slough, J.~Prager, and R.~Winglee.
\newblock High power helicon thruster.
\newblock In {\em 41st AIAA/ASME/SAE/ASEE Joint Propulsion Conference \&
  Exhibit}, Tucson, Arizona, 2005.

\bibitem{Charles:2009aa}
C.~Charles.
\newblock Plasmas for spacecraft propulsion.
\newblock {\em Journal of Physics D: Applied Physics}, 42(16):163001, 2009.

\bibitem{Batishchev:2009aa}
O.~V. Batishchev.
\newblock Minihelicon plasma thruster.
\newblock {\em Plasma Science, IEEE Transactions on}, 37(8):1563, 2009.

\bibitem{Loewenhardt:1991aa}
P.~K. Loewenhardt, B.~D. Blackwell, R.~W. Boswell, G.~D. Conway, and S.~M.
  Hamberger.
\newblock Plasma production in a toroidal heliac by helicon waves.
\newblock {\em Phys. Rev. Lett.}, 67:2792, 1991.

\bibitem{Hanna:2001aa}
J.~Hanna and C.~Watts.
\newblock Alfv{\'e}n wave propagation in a helicon plasma.
\newblock {\em Physics of Plasmas}, 8(9):4251, 2001.

\bibitem{Petrzilka:1994aa}
V.~Petrzilka and J.~A. Tataronis.
\newblock Non-resonant currents driven by helicon waves.
\newblock {\em Plasma Physics and Controlled Fusion}, 36(6):1027, 1994.

\bibitem{Zhu:1989aa}
P.~Zhu and R.~W. Boswell.
\newblock Ar ii laser generated by {L}andau damping of whistler waves at the
  lower hybrid frequency.
\newblock {\em Phys. Rev. Lett.}, 63:2805, 1989.

\bibitem{Chen:1996aa}
F.~F Chen, I.~D. Sudit, and M.~Light.
\newblock Downstream physics of the helicon discharge.
\newblock {\em Plasma Sources Science and Technology}, 5(2):173, 1996.

\bibitem{Shamrai:1996aa}
K.~P. Shamrai and V.~B. Taranov.
\newblock Volume and surface rf power absorption in a helicon plasma source.
\newblock {\em Plasma Sources Science and Technology}, 5(3):474, 1996.

\bibitem{Breizman:2000aa}
B.~N. Breizman and A.~V. Arefiev.
\newblock Radially localized helicon modes in nonuniform plasma.
\newblock {\em Phys. Rev. Lett.}, 84:3863, 2000.

\bibitem{Chang:2016ab}
L.~Chang, Q.~C. Li, H.~J. Zhang, Y.~H. Li, Y.~Wu, B.~L. Zhang, and Z.~Zhuang.
\newblock Effect of radial density configuration on wave field and energy flow
  in axially uniform helicon plasma.
\newblock {\em Plasma Science and Technology}, 18(8):848, 2016.

\bibitem{Klozenberg:1965aa}
J.~P. Klozenberg, B.~McNamara, and P.~C. Thonemann.
\newblock The dispersion and attenuation of helicon waves in a uniform
  cylindrical plasma.
\newblock {\em Journal of Fluid Mechanics}, 21:545, 1965.

\bibitem{Sudan:1967aa}
R.~N. Sudan, A.~Cavaliere, and M.~N. Rosenbluth.
\newblock Nonlinear interaction of helicons (whistlers) in inhomogeneous media.
\newblock {\em Phys. Rev.}, 158:387, 1967.

\bibitem{Chen:1997ab}
F.~F. Chen and D.~Arnush.
\newblock Generalized theory of helicon waves. i. normal modes.
\newblock {\em Physics of Plasmas}, 4(9):3411, 1997.

\bibitem{Arnush:1997aa}
D.~Arnush and F.~F. Chen.
\newblock Generalized theory of helicon waves. ii. excitation and absorption.
\newblock {\em Physics of Plasmas}, 5(5):1239, 1997.

\bibitem{Arnush:2000aa}
D.~Arnush.
\newblock The role of trivelpiece--gould waves in antenna coupling to helicon
  waves.
\newblock {\em Physics of Plasmas}, 7(7):3042, 2000.

\bibitem{Boswell:1984ab}
R.~W. Boswell.
\newblock Very efficient plasma generation by whistler waves near the lower
  hybrid frequency.
\newblock {\em Plasma Physics and Controlled Fusion}, 26(10):1147, 1984.

\bibitem{Boswell:1987aa}
R.~W. Boswell and R.~K. Porteous.
\newblock Large volume, high density rf inductively coupled plasma.
\newblock {\em Applied Physics Letters}, 50(17):1130, 1987.

\bibitem{Blackwell:2012aa}
B.~D. Blackwell, J.~F. Caneses, C.~M. Samuell, J.~Wach, J.~Howard, and C.~Corr.
\newblock Design and characterization of the magnetized plasma interaction
  experiment (magpie): a new source for plasma--material interaction studies.
\newblock {\em Plasma Sources Science and Technology}, 21(5):055033, 2012.

\bibitem{Squire:2006aa}
J.~P. Squire, F.~R. Chang-Diaz, T.~W. Glover, V.~T. Jacobson, G.~E. McCaskill,
  D.~S. Winter, F.~W. Baity, M.~D. Carter, , and R.~H. Goulding.
\newblock High power light gas helicon plasma source for vasimr.
\newblock {\em Thin Solid Films}, 506-507 (Supplement C):579--582, 2006.

\bibitem{Mori:2004aa}
Y.~Mori, H.~Nakashima, F.~W. Baity, R.~H. Goulding, M.~D. Carter, and D.~O.
  Sparks.
\newblock High density hydrogen helicon plasma in a non-uniform magnetic field.
\newblock {\em Plasma Sources Science and Technology}, 13(3):424, 2004.

\end{thebibliography}

\end{document}